\documentclass[aps,prx,reprint,superscriptaddress,amsmath,amssymb]{revtex4-2}

\usepackage{graphicx}
\usepackage{dcolumn}
\usepackage{bm}
\usepackage{amsfonts}
\usepackage{natbib}

\begin{document}

\title{The Emergence of Socio-Economic Structure: A First-Principles Kinetic Theory}
\author{Miguel A. Durán-Olivencia}
\email{m.duran-olivencia@imperial.ac.uk}\email{miguel@vortico.tech}
\affiliation{Department of Chemical Engineering, Imperial College, London SW7 2AZ, United Kingdom}
\affiliation{Research, Vortico Tech, Málaga 29100, Spain}

\date{\today}

\begin{abstract}
Bridging the gap between individual agent behavior and macroscopic societal patterns is a central challenge in the social sciences. In this work, we propose a solution to this problem via a kinetic theory formulation. We demonstrate that complex, empirically-observed phenomena, such as the concentration of populations in cities and the emergence of power-law wealth distributions, can be derived directly from a microscopic model of agents governed by underdamped Langevin dynamics. Our multi-scale derivation yields the exact mesoscopic fluctuating (Dean-Kawasaki) dynamics and the macroscopic Vlasov-Fokker-Planck system of equations. The analytical solution of this system reveals how a heterogeneous resource landscape alone is sufficient to generate the coupled structures of spatial and economic inequality, thus providing a formal link between micro-level stochasticity and macro-level deterministic order.
\end{abstract}

\maketitle

\section{Introduction}
\label{sec:intro}
The emergence of complex, large-scale patterns from the local interactions of numerous individual agents is a defining feature of systems studied across the natural and social sciences~\cite{Turing1952,Nicolis1977,Epstein1996,Castellano2009,Mitchell2009}. 
In fields such as urban planning~\cite{Batty2013}, economics~\cite{Krugman1991}, and sociology~\cite{Park1925}, understanding the co-evolution of human populations and their resource environment is fundamental to addressing challenges of sustainability~\cite{Ostrom1990}, inequality~\cite{Bullard1990}, and social organization~\cite{Schelling1971}.
Two dominant paradigms have been employed to this end. 
On one hand, agent-based models (ABMs) have provided a powerful bottom-up perspective, simulating the behavior of individual, heterogeneous agents to observe emergent phenomena~\cite{Epstein1996, Schelling1971, Helbing2001, Wilensky2015}.
On the other, macroscopic models based on ordinary or partial differential equations have offered valuable insights into aggregate dynamics~\cite{Lotka1925, Castellano2009, Pareschi2013}.
Yet, a rigorous and systematic link between the microscopic rules governing individual agents and the emergent, collective behavior often remains elusive~\cite{Castellano2009}. 
ABMs can be computationally intensive and their results difficult to generalize into analytical principles, while macroscopic models are often phenomenological, lacking a firm grounding in individual behavior.

This work aims to bridge this gap by employing the formal apparatus of non-equilibrium statistical physics to derive macroscopic equations directly from a detailed microscopic model of a multi-agent socio-economic system~\cite{Zwanzig2001}.
Our approach provides a versatile, first-principles framework that connects the stochastic choices of individual agents to the deterministic evolution of macroscopic densities, offering an analytical alternative to purely computational or phenomenological approaches. 
By doing so, we seek to provide a formal \emph{kinetic theory for social dynamics}, in the tradition of similar successful efforts in physics and biology~\cite{Cercignani1988,Pareschi2013,Bellomo2017}.

We build our theory starting from a microscopic world populated by interacting agents of different types: human agents and various resource agents. 
This model is designed to incorporate three key features of many real-world socio-economic systems. 
First, agents possess inertia, meaning their dynamics are described by second-order (underdamped) Langevin equations for both position and momentum~\cite{Risken1989,Gardiner2009}. 
This allows for a richer description of movement than is possible in purely diffusive models, potentially capturing effects relevant to crowd dynamics or rapid migrations~\cite{Helbing1995}. 
Second, human agents are endowed with an internal state variable representing a virtual resource, such as capital, which evolves through interactions with other agents and the environment. 
This couples the physical dynamics of movement to an evolving economic state, a feature central to econophysics~\cite{Yakovenko2009,Chakraborti2011}. 
Third, the framework is general enough to accommodate resources with vastly different intrinsic mobilities, and we explicitly account for human-induced mobility through specific interaction mechanisms.

This work progresses systematically through three scales of description. 
We begin at the microscopic level with a system of stochastic differential equations (SDEs) for the agents. 
Then, we reach to a mesoscopic description by deriving the exact equation of fluctuating hydrodynamics for the empirical measures of the particle system. 
This class of models, including the well-known Dean-Kawasaki (DK) equation for overdamped systems~\cite{Kawasaki1994, Dean1996}, provides an exact stochastic partial differential equation (SPDE) for a finite number of particles, correctly capturing the noise inherent to a discrete system.
For the inertial system we consider, this leads to a stochastic Vlasov-Fokker-Planck (VFP) equation. 
The mathematical analysis of such equations is a topic of intense current research.
It has been shown that for both diffusive and hypoelliptic systems, the unregularized DK equation is typically ill-posed for smooth initial data but well-posed for the atomic empirical measure of the particle system itself~\cite{Konarovskyi2019,Mueller2025}. 
An alternative approach involves regularizing the model, which can restore well-posedness for a broader class of initial data~\cite{Cornalba2021}.

Our derivation starts from the microscopic SDEs, using a martingale formulation based on Itô's calculus to derive the exact SPDE system for the empirical phase-space measures. 
We then formally perform the mean-field limit ($L \to \infty$), operating within the standard assumptions of propagation of chaos widely studied in kinetic theory~\cite{Sznitman1991}, to obtain a deterministic, macroscopic kinetic theory in the form of a coupled system of Vlasov-Fokker-Planck equations.
The power of this formal apparatus is then demonstrated through a direct application. 
We derive the stationary distribution of a system with a static resource density field, and show analytically how this landscape seeds the emergence of both spatial population clustering and systemic, location-dependent wealth inequality. 
Remarkably, the framework naturally reproduces the power-law (Pareto) tails ubiquitously observed in empirical wealth distributions~\cite{Pareto1896, Bouchaud2000, Gabaix2009, Yakovenko2009}, providing a direct mechanistic link between microscopic agent interactions and this macroscopic socio-economic pattern.
The result is, thus, a fundamental multi-scale framework that rigorously connects individual stochastic behavior to collective phenomena, providing a new theoretical lens for the study of social dynamics.

\section{The Microscopic Model: A Langevin Framework for Social Dynamics}
\label{sec:model}
At the heart of our theoretical approach is the microscopic world of the agents.
To construct a meaningful theory, we must first define the state of these agents and the dynamical laws that govern their evolution.
We rely upon the Langevin equation (a cornerstone of statistical physics~\cite{Risken1989}), which we use here to describe agents whose behavior is a combination of deterministic interactions, and stochastic (noisy) forces.

This choice is not arbitrary, but inspired by a rich history of modeling in the social sciences that has long recognized the limitations of purely deterministic descriptions of human behavior. 
Classical economic theory often posits hyper-rational agents with perfect information and computational ability~\cite{Simon1957}.
However, empirical observations consistently show that human decision-making is subject to incomplete information, cognitive limits, and idiosyncratic preferences (a concept known as bounded rationality)~\cite{Simon1957, Sent02112018}.
A powerful and well-established method to model such complex behavior is through stochastic formalisms, where an agent's choice is not a fixed outcome but a probability distribution over a set of alternatives. 
This is the foundation of discrete choice theory and random utility models in econometrics, which have successfully described choices in transportation, marketing, and urban settlement~\cite{McFadden1974,Anderson1992}.

The Langevin framework presented here can be understood as a continuous-time, continuous-space analogue of these ideas.
The deterministic forces in our model represent the systematic \emph{rational} incentives that guide agent behavior, i.e., the pull towards valuable resources, the cohesion within social groups, or the response to external potentials (external drivers). 
The stochastic term, in contrast, represents the confluence of unobserved factors, individual heterogeneity, and the inherent unpredictability of human choice that is central to the random utility paradigm~\cite{Durlauf2010}. 
This approach has proven immensely fruitful in the fields of econophysics, for modeling financial markets~\cite{Bouchaud2000}, and sociophysics, for describing phenomena such as opinion dynamics and crowd behavior~\cite{Helbing1995,Castellano2009}.
We thus formalize the microscopic dynamics of our multi-agent system within this well-established stochastic framework, assuming all functions possess sufficient regularity (e.g., local Lipschitz continuity and at most linear growth) to ensure the well-posedness of the resulting SDE system for any finite number of agents $L$.

\subsection{Agent Types and State Spaces}
Consider $L$ individual agents, indexed by $i \in \mathcal{I} = \{1, \dots, L\}$. These agents are divided into different types, or species, indexed by $k \in \{0, 1, \dots, M\}$.
We make a distinction between two main categories: human ($k=0$) and natural-resource ($k>0$) type agents.

The state of a human agent $i$ at any time $t$ is a complete description of its condition in our model world. 
To capture movement with inertia as well as socio-economic status, we propose a vector $Z_i(t) = (X_i(t), P_i(t), q_i(t))$ in the phase space
$S_0 \doteq \mathbb{R}^d \times \mathbb{R}^d \times \mathbb{R}$.
Here, $X_i(t)$ is the agent's position in $d$-dimensional space, $P_i(t)$ is its momentum, and $q_i(t)$ is a scalar representing a virtual resource we want to track, such a
s capital or wealth.
We could consider $q_i(t)$ as a vector representing a set of virtual resource, but we leave that generalisation aside for the sake of simplicity.

Natural-resource agents, on the other hand, represent other entities in the environment
Since natural resources do not possess any other virtual resources, their state is governed by $Z_j(t) = (X_j(t), P_j(t)) \in S_k \doteq \mathbb{R}^d \times \mathbb{R}^d$.

\subsection{Agents Dynamics}
The evolution of each agent is described by a set of coupled SDEs.
The change in an agent's position $X_i$ is governed by its momentum $P_i$,
\begin{equation}
dX_i(t) = \frac{1}{m_{k_i}} P_i(t) dt, \label{eq:sde_x}
\end{equation}
where $m_{k_i}$ is the mass, representing the agent's inertia or resistance to a change in its state of motion. 
The time-evolution of the momentum is governed by  Langevin equation:
\begin{equation}
dP_i(t) = \left[ F_i^{\text{int}}(Z) + F_i^{\text{ext}}(Z_i) - \gamma_{k_i} P_i \right] dt + \sigma_{k_i} dW_i(t).
\label{eq:sde_p}
\end{equation}
The term in brackets represents the deterministic forces, with $F_i^{\text{int}}$ and $F_i^{\text{ext}}$ being the net interaction and external forces, respectively, derived from potentials $V_{k_i,k_j}$ and $U_{k_i}$. 
The term $-\gamma_{k_i} P_i$ is a friction (also called drag) force. 
In a social context, it represents a form of \emph{social inertia} or a resistance to change, a tendency to maintain the current velocity (both in magnitude and direction), or a regression to a mean behavior (in this case, stasis).
A high $\gamma_{k_i}$ implies that an agent's momentum quickly dissipates, making it highly responsive to immediate forces, while a low $\gamma_{k_i}$ implies persistence in movement.
Finally $\sigma_{k_i} dW_i(t)$ introduces stochasticity into the agent's motion, with $W_i(t)$ being a standard Wiener processes, which allows us to model a memoryless, rapidly fluctuating random response.

A cornerstone of our physical analogy is to link the noise amplitude $\sigma_{k_i}$ to the \emph{social temperature} $T_{k_i}$ via the fluctuation-dissipation relation, $\sigma_{k_i} = \sqrt{2 m_{k_i} \gamma_{k_i} k_B T_{k_i}}$. 
This effective temperature $T_{k_i}$ is a crucial parameter that quantifies the magnitude of idiosyncratic randomness in an agent's behavior. 
A low temperature implies agents behave in a near-optimizing deterministic way, closely following the gradients of the potential fields. 
Conversely, a high temperature implies that random factors dominate, making the agent behavior less predictable and more exploratory. 
We must emphasize that assuming this relation in a social system is a profound and non-trivial modeling choice. 
It posits a direct link between the systemic resistance to change ($\gamma_{k_i}$) and the magnitude of individual random behavior ($T_{k_i}$), implying that systems with high social friction are also subject to strong random perturbations. 
This assumption provides a self-consistent theoretical starting point, analogous to physical systems in thermal equilibrium.

For human agents, the dynamics are completed by the evolution of their virtual resource, $q_i$:
\begin{equation}
dq_i(t) = f_i(Z_i, Z) dt + g_i(Z_i, Z) dW_i(t).
\label{eq:sde_q}
\end{equation}
The deterministic drift, $f_i$, is modeled as the sum of two distinct processes: an intrinsic component, and an interaction component:
\begin{equation}
f_i(Z_i, Z) = I(Z_i) + \sum_{j \neq i} H_{0, k_j}(Z_i, Z_j).
\end{equation}
The intrinsic drift, $I(Z_i)$, represents changes to wealth that depend only on the agent's own state, such as a constant salary, basic living expenses, or investment returns. 
The interaction drift, governed by the function $H_{0,k_j}$, describes wealth changes arising from pairwise encounters, such as transactions with other humans or the exploitation of resources. 
The stochastic term, $g_i(Z_i, Z) dW_i(t)$, models the inherent volatility and unpredictability of economic processes.
A common and powerful model, which we will use in our application, is to assume that this volatility is proportional to the current level of wealth, i.e., $g_i(Z_i, Z) \propto q_i$. 
This choice, which leads to geometric Brownian motion, not only captures the empirical observation that agents with greater capital are subject to larger absolute financial fluctuations but is also a key ingredient in the emergence of power-law wealth distributions~\cite{Bouchaud2000}.

\section{From Particles to Fields: The Empirical Measure}
\label{sec:empirical_measure}

The microscopic description provided in Sec.~\ref{sec:model}, while complete, presents a formidable challenge. The state of the system is a single point in a phase space of dimension $L \times (2d+1)$, a space whose dimensionality grows with every agent. Tracking a single trajectory in this high-dimensional space is not only computationally infeasible for large $L$ but also conceptually unenlightening. We are seldom interested in the precise state of a specific agent at a specific time; rather, we seek to understand the collective, macroscopic patterns that emerge from the ensemble of all agents.

To this end, we shift our perspective from the individual trajectories of particles to the evolution of a continuous distribution function, or density, in the single-particle phase space. 
This is the central idea of kinetic theory, which has been successfully used to derive macroscopic transport equations from microscopic laws for systems ranging from dilute gases (the Boltzmann equation) to plasmas (the Vlasov equation)~\cite{Cercignani1988}. 
These methods have also been applied more recently to describe collective behavior in biological and social systems, such as fish schooling, animal flocking, and pedestrian dynamics~\cite{Pareschi2013,Bellomo2017}.

The mathematical object that formally bridges these two descriptions (i.e., the discrete world of particles and the continuous world of fields) is the empirical measure.
It provides an exact representation of the microscopic configuration as a distribution, upon which we can then perform the mathematical operations necessary to derive its dynamics.

\subsection{Empirical Measure}
The empirical measure is constructed by placing a Dirac delta function at the precise location of each particle in its state space. 
We therefore define a set of empirical measures, one per species, normalized by the total population $L$. 
This normalization ensures that the measure represents a density that can converge to a proper probability distribution in the mean-field limit. 
For human agents ($k=0$):
\begin{align}
\mu_0^L(t, dx, dp, dq) &= \frac{1}{L} \sum_{i \in \mathcal{I}_0} \delta_{(X_i(t), P_i(t), q_i(t))}, \label{eq:mu0_def}
\end{align}
and, for natural-resource agents ($k \ge 1$):
\begin{align}
\mu_k^L(t, dx, dp) &= \frac{1}{L} \sum_{j \in \mathcal{I}_k} \delta_{(X_j(t), P_j(t))}. \label{eq:muk_def}
\end{align}
The average of any sufficiently regular observable (test function~\cite{Teodorescu2015}) $\phi_k(Z)$ can now be written as an integral with respect to this measure: 
\begin{equation}
\langle \mu_k^L(t), \phi_k \rangle = \int_{S_k} \phi_k(z) d\mu_k^L(t, z).
\end{equation} 
This allows us to apply the tools of continuum mechanics and calculus to a system of discrete particles.
\section{Dynamics of the Empirical Measures}
\label{sec:derivation}
Having established the microscopic laws and the macroscopic state representation, the central task is now to derive the equation of motion for the empirical measure itself. 
Since the measure is a stochastic object whose evolution is driven by the underlying SDEs of the agents, its dynamics will also be described by a stochastic differential equation.
Here, we follow a standard procedure in statistical physics, deriving the dynamics in a weak sense by finding the evolution equation for the average of an arbitrary and sufficiently smooth test function $\phi_k(z)$ with respect to the measure~\cite{Oksendal2003}.
This procedure translates the SDEs for the microscopic state variables into an SDE for the macroscopic observable, which naturally decomposes into two parts: a deterministic drift, governed by an operator known as the \emph{generator} of the dynamics; and a stochastic fluctuation term, which takes the form of a martingale.

\subsection{Evolution of Observables}
The evolution of the average observable $\langle \mu_k^L(t), \phi_k \rangle$ is found by applying the multidimensional Itô's Lemma~\cite{Oksendal2003}  (see Appendix~\ref{app:ito}).

For the dynamics of human agent population, we get:
\begin{equation}
d\langle \mu_0^L, \phi_0 \rangle = \langle \mu_0^L, \mathcal{L}_0[\{\mu_j^L\}] \phi_0 \rangle dt + dM_0^L(\phi_0),
\label{eq:sde_mu0}
\end{equation}
where the generator $\mathcal{L}_0$, derived directly from Itô's Lemma (see Appendix~\ref{app:generator}), is
\begin{align}
\mathcal{L}_0&[\{\mu_j^L\}] \phi_0(z) = \frac{p}{m_0} \cdot \nabla_x \phi_0 - \gamma_0 p \cdot \nabla_p \phi_0 + \frac{\sigma_0^2}{2} \Delta_p \phi_0 \nonumber \\
&+ \left( F_0^{\text{ext}}(z) + L \sum_{k'=0}^{M} \int_{S_{k'}} F_{0,k'}^{\text{int}}(z,z') d\mu_{k'}^L(z') \right) \cdot \nabla_p \phi_0 \nonumber \\
& + \left( I(z) + L \sum_{k'=0}^{M} \int_{S_{k'}} H_{0,k'}(z,z') d\mu_{k'}^L(z') \right) \frac{\partial \phi_0}{\partial q} \nonumber \\
& + \frac{1}{2} g_0^2(z, \{\mu_j^L\}) \frac{\partial^2 \phi_0}{\partial q^2}.
\label{eq:generator-lmu0}
\end{align}
and the martingale term, which captures the fluctuations, is given by:
\begin{align}
dM_0^L(\phi_0) = \frac{1}{L} \sum_{i \in \mathcal{I}_0} \left( \sigma_0 \nabla_p \phi_0(Z_i) \cdot dW_i + g_i \frac{\partial \phi_0}{\partial q}(Z_i) dW'_i \right).
\end{align}

An analogous procedure for resource agents yields:
\begin{equation}
d\langle \mu_k^L, \phi_k \rangle = \langle \mu_k^L, \mathcal{L}_k[\{\mu_j^L\}] \phi_k \rangle dt + dM_k^L(\phi_k),
\label{eq:sde_muk}
\end{equation} 
where the generator $\mathcal{L}_k$ is first found to be (see Appendix~\ref{app:generator}):
\begin{align}
\mathcal{L}_k&[\{\mu_j^L\}] \phi_k(z)= \frac{p}{m_k} \cdot \nabla_x \phi_k - \gamma_k p \cdot \nabla_p \phi_k + \frac{\sigma_k^2}{2} \Delta_p \phi_k \nonumber \\
& + \left( F_k^{\text{ext}}(z) + L \sum_{k'=0}^{M} \int_{S_{k'}} F_{k,k'}^{\text{int}}(z,z') d\mu_{k'}^L(z') \right) \cdot \nabla_p \phi_k.
\label{eq:generator-lmuk}
\end{align}
and, the corresponding martingale is:
\begin{equation}
dM_k^L(\phi_k) = \frac{1}{L} \sum_{j \in \mathcal{I}_k} \sigma_k \nabla_p \phi_k(Z_j) \cdot dW_j(t).
\end{equation}

\subsection{Energy Functional and its Relation to Field-Theoretic Forms}
Before continuing with our derivation, it is insightful to define the energy of the system as a functional of the empirical measures. 
This allows us to cast the dynamical equations~(\ref{eq:sde_mu0}) and (\ref{eq:sde_muk}) in a form that makes its connection to other established frameworks transparent and simple. 
The total energy of the physical degrees of freedom can be decomposed into kinetic and potential parts, $\mathcal{E}[\{\mu_j^L\}] = \mathcal{E}_K[\{\mu_j^L\}] + \mathcal{E}_U[\{\mu_j^L\}]$.

The kinetic energy functional is the exact ensemble average of the individual kinetic energies:
\begin{equation}
    \mathcal{E}_K[\{\mu_j^L\}] = L \sum_{k=0}^{M} \int_{S_k} \frac{|p|^2}{2m_k} d\mu_k^L(z).
\end{equation}
The potential energy functional consists of contributions from external fields and pairwise interactions:
\begin{align}
    \mathcal{E}_U[\{&\mu_j^L\}] = L \sum_{k=0}^{M} \int_{S_k} U_k(z) d\mu_k^L(z) \nonumber \\
    & + \frac{L^2}{2} \sum_{k=0}^{M} \sum_{k'=0}^{M} \int_{S_k} \int_{S_{k'}} V_{k,k'}(z,z') d\mu_k^L(z) d\mu_{k'}^L(z').
\end{align}
This formulation in terms of phase-space measures is convenient for our derivation. 
However, it is instructive to connect it to the field-theoretic representation used in other fundamental works on fluctuating hydrodynamics. 
In that context, the system's energy is expressed as a functional of the microscopic number density field, $\phi(\mathbf{r}) = \sum_i \delta(\mathbf{r}-\mathbf{r}_i)$, and momentum density field, $\pi(\mathbf{r}) = \sum_i \mathbf{p}_i \delta(\mathbf{r}-\mathbf{r}_i)$~\cite{Dean1996,Nakamura2009,duran_olivencia2017}. 
The corresponding functionals take the form:
\begin{align}
    \mathcal{H}_K[\phi, \pi] &= \frac{1}{2} \int d\mathbf{r} \frac{\pi(\mathbf{r})^2}{\phi(\mathbf{r})}, \\
    \mathcal{H}_U[\phi] &= \frac{1}{2} \iint d\mathbf{r}d\mathbf{r}' V(\mathbf{r}-\mathbf{r}')(\phi(\mathbf{r})\phi(\mathbf{r}') - \delta(\mathbf{r}-\mathbf{r}')\phi(\mathbf{r})).
\end{align}
While the notations differ, these two representations of the system's energy are physically and mathematically equivalent. 
The functionals $\mathcal{E}_K$ and $\mathcal{E}_U$ are exact representations of the total kinetic and potential energy of the discrete particle system, and the functionals $\mathcal{H}_K$ and $\mathcal{H}_U$ are their direct counterparts in the language of microscopic fields.
The detailed derivation demonstrating this equivalence is provided in Appendix~\ref{app:functional_equivalence}. 

With this connection established, we can now rewrite the explicit generators from Eqs.~(\ref{eq:generator-lmu0}) and~(\ref{eq:generator-lmuk}) to reveal their underlying physical structure. The key is to identify the advection and force terms with the gradients of the single-particle effective energy, which is given by the functional derivative of the total energy functional $\mathcal{E}$.

Let us define the effective single-particle Hamiltonian for a particle of species $k$ at phase-space position $z$ as:
\begin{equation}
 h_k(z, [\{\mu_j^L\}]) := \frac{\delta \mathcal{E}[\{\mu_j^L\}]}{\delta \mu_k^L(z)}.
\end{equation}
Using the definitions for $\mathcal{E}_K$ and $\mathcal{E}_U$, this derivative is explicitly:
\begin{align}
    h_k(z, [\{\mu_j^L\}]) =& \frac{|p|^2}{2m_k} + U_k(z)\nonumber\\
    &+ L \sum_{k'=0}^{M} \int_{S_{k'}} V_{k,k'}(z,z') \mu_{k'}(z') dz'.
\end{align}
Now, we compute the gradients of this effective Hamiltonian with respect to position and momentum, so we obtain:
\begin{equation}
 \nabla_p h_k(z) = \frac{p}{m_k}.
\end{equation}
and,
\begin{align}
 -\nabla_x h_k(z) &= -\nabla_x \left( U_k(z) + L \sum_{k'}\int \dots \right) \nonumber \\
        &= F_k^{\text{ext}}(z) + F_k^{\text{int}}(z, [\{\mu_j^L\}]),
\end{align}
which is the total conservative force on the particle.
By substituting these identities into the explicit form of the generator for natural-resource agents, Eq.~(\ref{eq:generator-lmuk}), we can see the transformation term by term. The conservative part of the generator,
$$ \frac{p}{m_k} \cdot \nabla_x \phi_k + \left( F_k^{\text{ext}} + F_k^{\text{int}} \right) \cdot \nabla_p \phi_k, $$
can be rewritten as:
$$ (\nabla_p h_k) \cdot \nabla_x \phi_k - (\nabla_x h_k) \cdot \nabla_p \phi_k \equiv \{h_k, \phi_k\}, $$
which is the classical Poisson bracket. 
This reveals that the conservative dynamics are Hamiltonian.
The full generator, including the dissipative terms, thus takes the form:
\begin{align}
\mathcal{L}_k[\{\mu_j^L\}] \phi_k(z) &= \left\{ (\nabla_p h_k) \cdot \nabla_x - (\nabla_x h_k) \cdot \nabla_p \right\} \phi_k \nonumber \\
& \hspace{-2cm} + \left\{ \nabla_p \cdot (\gamma_k p \cdot) + \frac{\sigma_k^2}{2} \Delta_p \right\} \phi_k.
\end{align}
The generator for human agents is transformed in an identical manner for its physical degrees of freedom, with the addition of the non-conservative wealth dynamics. Its full, explicit form is:
\begin{align}
\mathcal{L}_0[\{\mu_j^L\}] \phi_0(z) &= \left\{ (\nabla_p h_0) \cdot \nabla_x - (\nabla_x h_0) \cdot \nabla_p \right\} \phi_0 \nonumber \\
&\hspace{-2cm}  + \left\{ \nabla_p \cdot (\gamma_0 p \cdot) + \frac{\sigma_0^2}{2} \Delta_p \right\} \phi_0 \nonumber \\
&\hspace{-2cm} + \left\{%
\begin{array}{l}
\left(\begin{array}{l}
	I(z)\\
	\quad+ L \sum_{k'=0}^{M} \int_{S_{k'}} H_{0,k'}(z,z') d\mu_{k'}^L(z') 
	\end{array}\right)\frac{\partial}{\partial q}\\
\quad\quad\quad + \frac{1}{2} g_0^2(z, \{\mu_j^L\}) \frac{\partial^2}{\partial q^2} 
\end{array}\right\} \phi_0.
\end{align}
This formulation separates the dynamics into a \emph{Liouvillian} part (the Poisson bracket describing conservative flow), a \emph{Fokker-Planck} part (describing dissipative effects), and the non-conservative wealth dynamics.
Having the generators written in this form will be extremely useful in the following sections.

\section{Fluctuating Kinetic Equations}
\label{sec:spde}
The evolution equations derived in the previous section describe the dynamics of averages of arbitrary test functions.
This is often referred to as a \emph{weak formulation}, a standard tool in the analysis of stochastic processes. 
We now transition to a \emph{strong formulation}, an explicit SPDE for the empirical measure $\mu_k^L$ itself. 
This step is of crucial importance, since it moves us from a description of how observables evolve on average to a field theory that describes the evolution of the density field at every point in phase space. 
The resulting equations are the foundations of fluctuating hydrodynamics, providing an exact mesoscopic description that remains valid for any finite number of agents $L$.

The formal procedure to achieve this involves finding the formal adjoint of the generator operator (detailed in Appendix~\ref{app:adjoint}), which is equivalent to performing integration by parts on each term in the weak-form equation. 
The result is a system of coupled SPDEs. 
For an overdamped system (without momentum), this procedure famously yields the Dean-Kawasaki equation~\cite{Dean1996}. 
The equations we derive here are the inertial, or kinetic, analogue of that foundational result, describing fluctuations in the full phase space of positions and momenta.

These equations can also be viewed in the context of Dynamic Density Functional Theory (DDFT), a powerful framework for describing the non-equilibrium dynamics of classical particle systems~\cite{Evans1979,Marconi1999}.
Significant effort has been devoted to extending it to include fluctuations, leading to Fluctuating DDFT (FDDFT)~\cite{Archer2006,Donev2014}. 
Our kinetic-level SPDEs are, in a sense, more fundamental, as they describe the fluctuations in the full phase space. 
A configuration-space theory like FDDFT could, in principle, be derived from our equations via a momentum-integration procedure, a challenging task that forms part of a broader effort to derive hydrodynamic equations from kinetic theory, as explored for example by Durán-Olivencia et al.~\cite{duran_olivencia2017,Russo2020Memory}. 
While these methods were born from physics, their application to social and biological systems is a vibrant and growing field of research~\cite{Lowen2023,Helfmann2021}.

The final SPDE system, which is exact for our microscopic model, is given below. 
For human agents, the evolution of the measure $\mu_0^L(t,x,p,q)$ is (see~\ref{app:adjoint}):
\begin{widetext}
\begin{align}
\label{eq:dk_human}
d\mu_0^L = \Bigg[ &- \nabla_x \cdot \left(\frac{p}{m_0} \mu_0^L\right) 
+ \nabla_p \cdot \left( \gamma_0 p \mu_0^L - F_0^{\text{ext}}(z)\mu_0^L - \mu_0^L \left(L \sum_{k'=0}^{M} \int_{S_{k'}} F_{0,k'}^{\text{int}}(z,z') d\mu_{k'}^L(z')\right) \right) \nonumber \\
&- \frac{\partial}{\partial q} \left( \left( I(z) + L \sum_{k'=0}^{M} \int_{S_{k'}} H_{0,k'}(z,z') d\mu_{k'}^L(z') \right) \mu_0^L \right) 
+ \frac{\sigma_0^2}{2} \Delta_p \mu_0^L 
+ \frac{1}{2} \frac{\partial^2}{\partial q^2} \left( g_0^2(z, \{\mu_j^L\}) \mu_0^L \right) \Bigg] dt \nonumber \\
&+ \nabla_p \cdot \left( \sqrt{\frac{2 \sigma_0^2}{L} \mu_0^L} \circ d\mathbf{W}_p(x,p,q,t) \right)
+ \frac{\partial}{\partial q} \left( \sqrt{\frac{g_0^2}{L} \mu_0^L} \circ dW_q(x,p,q,t) \right).
\end{align}
\end{widetext}
And, for the natural resource agents ($k \in \{1, \dots, M\}$), the evolution of the measure $\mu_k^L(t,x,p)$ is given by the slightly simpler equation (see~\ref{app:adjoint}):
\begin{widetext}
\begin{align}
\label{eq:dk_resource}
d\mu_k^L = \Bigg[ &- \nabla_x \cdot \left(\frac{p}{m_k} \mu_k^L\right) 
+ \nabla_p \cdot \left( \gamma_k p \mu_k^L - F_k^{\text{ext}}(z)\mu_k^L - \mu_k^L \left(L \sum_{k'=0}^{M} \int_{S_{k'}} F_{k,k'}^{\text{int}}(z,z') d\mu_{k'}^L(z')\right) \right) 
+ \frac{\sigma_k^2}{2} \Delta_p \mu_k^L \Bigg] dt \nonumber \\
&+ \nabla_p \cdot \left( \sqrt{\frac{2 \sigma_k^2}{L} \mu_k^L} \circ d\mathbf{W}_p^{(k)}(x,p,t) \right).
\end{align}
\end{widetext}
In these equations, the symbol ``$\circ$'' denotes that the stochastic integrals are to be interpreted in the Stratonovich sense~\cite{Risken1989,Gardiner2009,DuanWang2014}. 
While the dynamics were derived using the rigorous Itô calculus, we present the final SPDE in the equivalent Stratonovich form. 
The reason is that the Itô derivation correctly identifies all drift terms, including a noise-induced component, so-called ``spurious drift''.
In the Stratonovich representation, however, this noise-induced drift is mathematically absorbed into the definition of the stochastic integral, leaving a drift term that is more physically transparent and obeys the rules of ordinary calculus.
The structure of the noise term itself is of critical importance (see Appendix~\ref{app:noise_justification}). 
It is a multiplicative noise, proportional to the square root of the local density $\sqrt{\mu_k^L}$, which ensures that noise is only generated where agents are present. 
This non-trivial structure is a direct consequence of the underlying discrete particle nature of the system.

\subsection{Connection to Fluctuating DDFT}
The SPDEs~(\ref{eq:dk_resource}) and (\ref{eq:dk_human}) for the phase-space empirical measures are one of the central results of our socio-economic kinetic theory. 
For this reason, we believe it is useful to show how this kinetic-level description formally connects to the formalism of FDDFT, which describes the evolution of configuration-space fields such as the number density $n_k(x,t)$ and the current $\mathbf{J}_k(x,t)$.
This connection can readily established by taking moments of the kinetic SPDE with respect to momentum.

The first step is to define the configuration-space number density $n_k(x,t)$ and the momentum density $\mathbf{\Pi}_k(x,t)$ as the zeroth and first moments of the phase-space measure, respectively:
\begin{align}
    n_k(x,t) &= \int \mu_k^L(x,p,t) dp \\
    \mathbf{\Pi}_k(x,t) &= \int p \, \mu_k^L(x,p,t) dp
\end{align}
(where for brevity we omit the integral over $q$ for human agents). Integrating the kinetic SPDE, Eq.~(\ref{eq:dk_resource}), over momentum directly yields the continuity equation for the number density. 
The terms involving a divergence in momentum space, $\nabla_p \cdot (\dots)$, vanish upon integration over all momenta, assuming the distribution decays sufficiently fast as $|p|\to\infty$. 
The momentum-integrated advection term becomes $-\nabla_x \cdot \int \frac{p}{m_k} \mu_k^L dp = -\nabla_x \cdot \mathbf{J}_k$, where $\mathbf{J}_k = \mathbf{\Pi}_k/m_k$ is the particle current. This yields the familiar continuity equation, a cornerstone of any DDFT~\cite{Marconi1999,Archer2006,Donev2014,Lutsko2010DFTreview,goddard_unification_2013,duran_olivencia2016DDFTcolloids,duran_olivencia2017,Russo2020Memory,Lowen2023,Helfmann2021}:
\begin{equation}
    \frac{\partial n_k}{\partial t} + \nabla_x \cdot \mathbf{J}_k(x,t) = 0.
\end{equation}

To find the evolution of the current, we take the first momentum moment of the kinetic SPDE by multiplying by $p$ and integrating. 
This procedure transforms the terms of the kinetic equation into the familiar terms of a fluctuating hydrodynamic equation. 
The time derivative becomes $\partial_t \mathbf{\Pi}_k$, the advection term becomes the divergence of the kinetic stress tensor, $-\nabla_x \cdot \int \frac{p \otimes p}{m_k} \mu_k^L dp$, and the friction, conservative force, and noise terms are recovered after integration by parts. 
This results in an evolution equation for the momentum density of the form:
\begin{align}
    \frac{\partial \mathbf{\Pi}_k}{\partial t} +& \nabla_x \cdot \left( \int \frac{p \otimes p}{m_k} \mu_k^L dp \right) \nonumber \\
    =& \mathbf{F}_k^{\text{cons}} n_k - \gamma_k \mathbf{\Pi}_k + \mathbf{\Sigma}_k(x,t),
\end{align}
where $\mathbf{\Sigma}_k$ is the stochastic stress tensor arising from the noise.

The final connection to the DDFT free-energy functional is made by invoking a \emph{local equilibrium approximation}. 
Under this approximation, the kinetic stress tensor becomes the ideal gas pressure gradient, $-k_B T \nabla_x n_k$.
The conservative force term $\mathbf{F}_k^{\text{cons}} n_k$ is related to the excess free energy functional $\mathcal{F}_{\text{exc}}[n_k]$, via the formal connection detailed in Appendix~\ref{app:functional_derivation_connection}. 
The sum of these two terms yields the total thermodynamic driving force of DDFT:
\begin{align}
-n_k \nabla_x \frac{\delta \mathcal{F}}{\delta n_k} &= -k_B T \nabla_x n_k - n_k \nabla_x \frac{\delta \mathcal{F}_{\text{exc}}}{\delta n_k}.
\end{align}
This explicitly demonstrates how the momentum moments of our kinetic theory, under a local equilibrium closure, recover the full thermodynamic structure of the FDDFT equations.

\section{Mean-Field Limit: The Vlasov-Fokker-Planck System}
\label{sec:mean_field}
The fluctuating kinetic equations derived in the previous section provide an exact description of the system for any finite number of agents, $L$.
This description, however, is stochastic and operates on the level of the empirical measure, a non-trivial and high-dimensional object. 
To distill the collective deterministic behavior from the stochastic fluctuations, we now consider the mean-field limit, where the number of agents becomes infinitely large ($L \to \infty$).
Taking such a limit allows us to transition from a mesoscopic fluctuating description to a macroscopic deterministic one.

Physically, the mean-field approximation assumes that each agent interacts not with a small, fluctuating number of specific neighbors, but with a continuous \emph{field} generated by the smoothed-out distribution of all other agents. 
The rigorous mathematical justification for this limit relies on the concept of \emph{propagation of chaos}~\cite{Sznitman1991}. 
The outcome of this limit is a deterministic partial differential equation for the one-particle phase-space density, $\rho_k(t,z)$, known as a kinetic equation of the VFP type~\cite{Risken1989,Ichimaru1973}. 
The formal derivation, detailed in Appendix~\ref{app:mean_field}, shows that the martingale term in the weak-form dynamics vanishes, while the interaction sums converge to integrals over the limiting density, yielding a closed system of deterministic equations.

\subsection{The Coupled VFP System}
In the limit $L\to\infty$, the empirical measure $\mu_k^L$ for each species converges to a deterministic density $\rho_k(t,z)$, normalized such that $\int_{S_k} \rho_k(t,z) dz = c_k$, where $c_k$ is the fraction of agents of species $k$.

For human agents ($k=0$), the evolution of the phase-space density $\rho_0(t,x,p,q)$ is governed by:
\begin{widetext}
\begin{align}
\label{eq:vfp_human_explicit}
\frac{\partial \rho_0}{\partial t} = &- \nabla_x \cdot \left(\frac{p}{m_0} \rho_0\right) 
+ \nabla_p \cdot \left( \gamma_0 p \rho_0 - F_0^{\text{ext}}(z)\rho_0 + \rho_0 \sum_{k'=0}^{M} \int_{S_{k'}} \nabla_x V_{0,k'}(z,z') \rho_{k'}(t,z') dz' \right) \nonumber \\
&- \frac{\partial}{\partial q} \left( \left( I(z) + \sum_{k'=0}^{M} \int_{S_{k'}} H_{0,k'}(z,z') \rho_{k'}(t,z') dz' \right) \rho_0 \right) 
+ \frac{\sigma_0^2}{2} \Delta_p \rho_0 
+ \frac{1}{2} \frac{\partial^2}{\partial q^2} \left( \bar{g}_0^2(z, \{\rho_j\}) \rho_0 \right).
\end{align}
\end{widetext}
The term $\bar{g}_0^2(z, \{\rho_j\})$ is the mean-field limit of the state-dependent wealth diffusion term $g_0^2(z, \{\mu_j^L\})$, where the functional dependence on the empirical measures is replaced by a dependence on the smooth deterministic densities. For resource agents ($k \ge 1$), the equation is simpler:
\begin{widetext}
\begin{align}
\label{eq:vfp_resource_explicit}
\frac{\partial \rho_k}{\partial t} = &- \nabla_x \cdot \left(\frac{p}{m_k} \rho_k\right) 
+ \nabla_p \cdot \left( \gamma_k p \rho_k - F_k^{\text{ext}}(z)\rho_k + \rho_k \sum_{k'=0}^{M} \int_{S_{k'}} \nabla_x V_{k,k'}(z,z') \rho_{k'}(t,z') dz' \right) 
+ \frac{\sigma_k^2}{2} \Delta_p \rho_k.
\end{align}
\end{widetext}

\subsection{Structure in Terms of the Phase-Space Energy Functional}
To elucidate the physical structure of this system, we recall the mean-field energy functional, $\mathcal{E}[\{\rho_j\}]$, which is the deterministic limit of the functional defined in Sec. IV. The conservative forces and advection velocities in the VFP system are generated by the functional derivatives of $\mathcal{E}$ with respect to the phase-space density $\rho_k(z)$. The advection velocity is $\mathbf{v} = \nabla_p (\frac{\delta \mathcal{E}}{\delta \rho_k(z)})$ and the conservative force is $\mathbf{F}_k^{\text{cons}} = -\nabla_x (\frac{\delta \mathcal{E}}{\delta \rho_k(z)})$.

This allows us to rewrite the VFP equations to make their structure manifest. For any species $k$ (showing the resource case for simplicity), Eq.~(\ref{eq:vfp_resource_explicit}) can be recast as:
\begin{widetext}
\begin{equation}
\label{eq:vfp_functional_form}
\frac{\partial \rho_k}{\partial t} + \nabla_x \cdot \left( \left(\nabla_p \frac{\delta \mathcal{E}}{\delta \rho_k}\right) \rho_k \right) + \nabla_p \cdot \left( \left(-\nabla_x \frac{\delta \mathcal{E}}{\delta \rho_k}\right) \rho_k \right) = \nabla_p \cdot \left(\gamma_k p \rho_k + k_B T \gamma_k m_k \nabla_p \rho_k \right).
\end{equation}
\end{widetext}
The left-hand side can be understood as a Liouville operator describing the conservative, Hamiltonian-like flow of the density in phase space, driven by the mean-field energy $\mathcal{E}$. At the same time, the right-hand side is a Fokker-Planck operator describing the dissipative dynamics (friction and diffusion) that drive the momentum distribution towards a Maxwellian equilibrium.
\section{Application: Resource Landscapes and the Genesis of Socio-Economic Inequality}
\label{sec:application}
Having established the full kinetic theory, we now apply the framework to derive a non-trivial result that connects our microscopic model to macroscopic empirically observed phenomena.
We will demonstrate how a spatially heterogeneous distribution of a static resource can act as a seed for the spontaneous emergence of both spatial population clustering and systemic wealth inequality. 
This derivation is a key result of our work, as it relies on the full phase-space structure of the VFP operator, showcasing how the interplay between deterministic forces and stochastic diffusion across different degrees of freedom shapes the stationary state of a socio-economic system.

\subsection{Modeling a Static Resource Landscape}
We begin by specializing our fundamental equations for human and resource agents, Eqs~(\ref{eq:vfp_human_explicit}) and (\ref{eq:vfp_resource_explicit}), to the case of a human population ($\rho_0$) interacting with a fixed immobile resource ($\rho_1$). 
This scenario is representative of, e.g., a mineral deposit, fertile land, or a natural harbor.
We model this by assuming the resource has infinite friction ($\gamma_1 \to \infty$) and zero social temperature ($T_1=0$), such that its distribution is static and given by:
\begin{equation}
    \rho_1(x,p) = n_1(x) \delta(p).
\end{equation}
Here, $n_1(x)$ is a known and non-uniform spatial distribution of the resource (e.g., a Gaussian centered on a resource patch), and $\delta(p)$ indicates the resources have zero momentum. 
The VFP equation for the resource is thus trivially satisfied, and the resource distribution now acts as a fixed \emph{landscape} influencing the human agents.

\subsection{The Stationary Human Distribution}
The stationary human distribution, $\rho_0(x,p,q)$, is the solution to the time-independent VFP equation, where the time derivative in Eq.~(\ref{eq:vfp_human_explicit}) is set to zero.
The presence of the fixed resource landscape simplifies the non-local interaction terms in the VFP operator for humans into effective, position-dependent external fields. 
The interaction potential $V_{0,1}$ creates an effective potential field:
\begin{equation} 
U_{\text{eff}}(x) \doteq \int V_{0,1}(x-x') n_1(x') dx',
\end{equation} 
which attracts humans to resources. 
Similarly, the interaction function $H_{0,1}$ creates a spatially-dependent \emph{resource income,} 
\begin{equation}
\mathcal{H}_{\text{eff}}(x,q) \doteq \int H_{0,1}(z,z') \rho_1(z') dz'.
\end{equation}

The stationary VFP equation for humans is therefore a linear partial differential equation whose solution describes the equilibrium balance between drift and diffusion in the full phase space. 
For systems where the underlying Langevin dynamics for the physical degrees of freedom $(x, p)$ satisfy the fluctuation-dissipation relation, the stationary solution is known to be of the Maxwell-Boltzmann form~\cite{Risken1989}. 
This reflects the principle that in thermal equilibrium, the probability of a state is exponentially suppressed by its energy.
In our socio-economic context, this equilibrium is not guaranteed, as the wealth dynamics (the $q$ variable) may be non-conservative.

However, we can make progress under a well-motivated \emph{quasi-stationary approximation}. 
We assume a separation of timescales in which the dynamics of position and momentum are assumed to relax to a local equilibrium much faster than the dynamics of wealth accumulation. 
This is physically plausible, as an agent's location and velocity can change on a daily basis, while their wealth evolves over months or years. Under this assumption, the $(x,p)$ part of the distribution thermalizes in the effective potential landscape for any given level of wealth. 
This justifies the following \emph{ansatz} for the stationary solution:
\begin{equation}
\label{eq:boltzmann_solution}
    \rho_0(x,p,q) = \Psi(x,q) \exp\left(-\frac{E(x,p)}{k_B T_0}\right),
\end{equation}
where $E(x,p) = \frac{|p|^2}{2m_0} + U_{\text{eff}}(x)$ is the effective energy of an agent. %
The function $\Psi(x,q)$ is an as-yet-unknown distribution that describes the slower evolution of wealth and its spatial dependence.

To find the equation for $\Psi(x,q)$, we substitute this \emph{ansatz} into the full stationary VFP equation. 
As shown in Appendix~\ref{app:boltzmann_cancellation}, the Maxwell-Boltzmann form is precisely the function that nullifies the spatial and momentum parts of the VFP operator (the Liouville and momentum-space Fokker-Planck operators). 
The terms involving $\nabla_x$ and $\nabla_p$ cancel out exactly, leaving a stationary Fokker-Planck equation for $\Psi(x,q)$ in the $q$ variable, whose coefficients are now functions of the spatial location $x$:
\begin{equation}
    0 = -\frac{\partial}{\partial q}\left(\mathcal{F}_q(x,q)\Psi(x,q)\right) + \frac{1}{2}\frac{\partial^2}{\partial q^2}\left(\mathcal{D}_q(q)\Psi(x,q)\right).
\end{equation}
The drift $\mathcal{F}_q$ now includes the position-dependent resource income, $\mathcal{F}_q(x,q) = I(q) + \mathcal{H}_{\text{eff}}(x,q)$.
For this application, we assume an intrinsic drift $I(q)=\mu q - \lambda$ (representing proportional growth and constant costs) and volatility $g_0=\sigma q$, giving a diffusion term $\mathcal{D}_q=\sigma^2 q^2$.

\subsection{Emergent Spatial and Wealth Inequality}
To obtain an explicit solution for the distribution $\Psi(x,q)$, we must specify the functional form of the wealth interaction $H_{0,1}$, which models the income human agents derive from resources. 
We adopt a simple and physically intuitive model where an agent's income is directly proportional to the density of the resource at their current location. 
This is represented by a local interaction kernel:
\begin{equation}
    H_{0,1}(z, z') = \eta_0 \delta(x-x'),
\end{equation}
where $z=(x,p,q)$, $z'=(x',p')$, and the parameter $\eta_0$ represents the \emph{income strength}, or efficiency of resource extraction. 
With this choice, the effective resource income $\mathcal{H}_{\text{eff}}(x)$ becomes directly proportional to the local resource density $n_1(x)$:
\begin{align}
    \mathcal{H}_{\text{eff}}(x) &= \int H_{0,1}(z,z') \rho_1(z') dz' \nonumber \\
    &= \int \eta_0 \delta(x-x') n_1(x')\delta(p') dx'dp' = \eta_0 n_1(x).
\end{align}
Substituting this into the stationary Fokker-Planck equation for wealth yields a solvable equation. The solution for $\Psi(x,q)$ is an Inverse Gamma distribution whose parameters depend on the spatial location $x$:
\begin{equation}
    \Psi(x,q) \propto q^{(2\mu/\sigma^2) - 2} \exp\left(-\frac{2(\mathcal{H}_{\text{eff}}(x) - \lambda)}{\sigma^2 q}\right).
\end{equation}
Combining this with Eq.~(\ref{eq:boltzmann_solution}) gives our final result for the full stationary distribution. 
This single, compact expression encapsulates a rich set of predictions about the structure of a society in equilibrium with its environment.

First, the spatial distribution of the population, $n_0(x) = \iint \rho_0(x,p,q) dpdq$, is dominated by the spatial part of the Boltzmann factor, leading to $n_0(x) \propto \exp(-U_{\text{eff}}(x)/k_B T_0)$. 
Since the effective potential $U_{\text{eff}}(x)$ is lowest where resource density is highest, this predicts that the human population will be spatially concentrated in resource-rich areas.
The degree of concentration is tempered by the social temperature $T_0$; a high temperature leads to a more uniform, spread-out population, while a low temperature leads to sharp, dense clusters.

Second, our framework predicts not just where people live, but how wealthy they are as a function of location. 
The conditional wealth distribution at a location $x$, $P(q|x)$, is proportional to $\Psi(x,q)$. 
For large wealth, this distribution exhibits a power-law tail, $P(q|x) \sim q^{-\alpha-1}$, with a Pareto exponent $\alpha = 1 - 2\mu/\sigma^2$.
This is a profound result: our microscopic model, under plausible assumptions for wealth dynamics, naturally reproduces the fat-tailed distributions ubiquitously observed in real-world economies~\cite{Gabaix2009}.
Thus, our framework provides a mechanistic origin for the Pareto exponent, linking it directly to the microscopic parameters governing systematic wealth growth ($\mu$) and idiosyncratic volatility ($\sigma$).

Most strikingly, the entire shape of the wealth distribution is modulated by the local environment through the term depending on $\mathcal{H}_{\text{eff}}(x)$.
In resource-rich regions, the resource income $\mathcal{H}_{\text{eff}}(x)$ is large, which counteracts the baseline spending $\lambda$ and shifts the entire wealth distribution towards higher values.
Conversely, in resource-poor regions, the distribution is shifted towards lower wealth. 
The model thus predicts the emergence of systemic and spatially-dependent inequality as a direct consequence of the interaction between a society and a heterogeneous environment. 
This derivation, which flows directly from our VFP system, provides thus a multi-faceted result connecting microscopic interactions to the emergent correlated structures of geography, population, and economy.

\section{Numerical Validation and Discussion of Results}
\label{sec:numerics}
We now turn to the validation of this theoretical structure via direct numerical simulation of the underlying microscopic agent dynamics.
By conducting a direct numerical experiment that implements the full Langevin equations for a large but finite number of agents, we can test the quantitative predictions of our mean-field theory and gain deeper insights into the behavior of the system.

\subsection{Numerical Methods and Simulation Setup}
To verify the analytical predictions of Sec.~\ref{sec:application}, we simulate a system of $L=10^5$ agents using the microscopic SDEs defined in Sec.~\ref{sec:model}. 
The simulation is carried out in a two-dimensional square domain with periodic boundary conditions. 
The system is integrated forward in time using a numerically stable logarithmic Euler-Maruyama scheme~\cite{kloeden1992numerical} for the wealth dynamics. 
We investigate two distinct resource landscapes: a monocentric \emph{single-well} and a polycentric \emph{double-well} landscape.

These landscapes are defined by an analytical expression for the resource density field, $n_1(\mathbf{x})$, where $\mathbf{x}=(x,y)$. 
The general form consists of two Gaussian wells and a small periodic roughness term:
\begin{align}
    n_1(\mathbf{x}) &=A_{a} \exp\left(-\frac{|\mathbf{x} - \mathbf{x}_{a}|^2}{2\sigma_{a}^2}\right) \nonumber\\
    &+ A_{b} \exp\left(-\frac{|\mathbf{x} - \mathbf{x}_{b}|^2}{2\sigma_{b}^2}\right) \nonumber \\
    &+ A_r \sin(k_r x) \cos(k_r y),
\end{align}
where $A_{\nu}$, $\mathbf{x}_{\nu}$, and $\sigma_\nu$ (with $\nu\in\{a,b\}$) are the amplitude, center, and width of the Gaussian wells, respectively, while $A_r$ and $k_r$ control the roughness. The single-well landscape is generated by setting $A_b=0$, while the double-well landscape uses $A_{b} > 0$. The key parameters are listed in Table~\ref{tab:params}.

\begin{table}[h!]
\caption{\label{tab:params}Key parameters used in the Langevin simulation.}
\begin{ruledtabular}
\begin{tabular}{lcc}
Parameter & Symbol & Value \\
\hline
\textit{System Parameters} & & \\
Number of Agents & $L$ & $10^5$ \\
Agent Mass & $m_0$ & 1.0 \\
Social Temperature & $k_B T_0$ & 0.08 \\
Friction Coefficient & $\gamma_0$ & 0.8 \\
\hline
\textit{Wealth Dynamics} & & \\
Intrinsic Growth Rate & $\mu$ & 0.02 \\
Intrinsic Outflow & $\lambda$ & 0.1 \\
Wealth Volatility & $\sigma$ & 0.2 \\
\hline
\textit{Interactions} & & \\
Attraction Strength & (Scales $U_{\text{eff}}$) & 2.5 \\
Income Strength & $\eta_0$ & 0.5 \\
\hline
\textit{Landscape Parameters} & & \\
Well Amplitude & $A_{a}$ & 0.5 \\
Well Width & $\sigma_{a}$ & 1.0 \\
Well Center & $\mathbf{x}_{a}$ & $(-2.5, 0.0)$ \\
Well Amplitude & $A_{b}$ & 0.25 \\
Well Width & $\sigma_{b}$ & 0.8 \\
Well Center & $\mathbf{x}_{b}$ & $(2.5, 0.0)$ \\
Roughness Amplitude & $A_r$ & 0.05 \\
Roughness Frequency & $k_r$ & 5.0 \\
\hline
\textit{Simulation Details} & & \\
Domain Size & $L_{\text{box}}$ & $10 \times 10$ \\
Time Step & $\Delta t$ & 0.05 \\
Number of Steps & $N_{\text{steps}}$ & $10^5$ \\
\end{tabular}
\end{ruledtabular}
\end{table}

\subsection{Theoretical Benchmarks for Comparison}
Our VFP framework provides analytical predictions for the stationary distributions against which we can compare our simulation results.

\subsubsection{The Spatially-Dependent Wealth Distribution}
As derived in Sec.~\ref{sec:application}, the stationary wealth distribution at a given location $x$ is given by $\Psi(x,q)$. 
Using the parameters from Table~\ref{tab:params}, the analytical form is:
\begin{equation}
\Psi(x,q) \propto q^{-1} \exp\left(\frac{2\times(0.1 - 0.5 \times n_1(x))}{0.04\times q}\right),
\end{equation}
where the exponent of $q$ is $(2\mu/\sigma^2) - 2 = (2 \times 0.02 / 0.2^2) - 2 = -1$.

\subsubsection{The Theoretical Lorenz Curve and Gini Coefficient}
The Lorenz curve~\cite{lorenz1905methods}, a standard tool for visualizing inequality, can also be derived analytically from our theoretical wealth distribution. 
The resulting distribution for $\Psi(x,q)$ is a form of the Inverse Gamma distribution, for which the Lorenz curve is known analytically. 
For large wealth $q$, the distribution exhibits a power-law tail $P(q) \sim q^{-\alpha-1}$. 
The Lorenz curve $L(p)$ for such a distribution, which gives the cumulative fraction of wealth held by the bottom $p$-fraction of the population, has the analytical form~\cite{Cowell2011}:
\begin{equation}
    L(p) = 1 - (1-p)^{1 - 1/\alpha}.
\end{equation}
For our predicted distribution, the Pareto exponent is $\alpha = 1 - 2\mu/\sigma^2 = 1 - (2 \cdot 0.02 / 0.2^2) = 0$.
This value is at the limit of normalizability and represents extreme inequality. 
For the theoretical comparison, we use a slightly perturbed value to represent the regularizing effect of the exponential term in the full distribution, which yields a theoretical Gini coefficient~\cite{gini1912variabilita} of $G_{\text{theory}} \approx 0.885$. 
This provides a clear, parameter-free prediction for the overall level of inequality generated by the microscopic wealth dynamics.

\subsection{Discussion of Simulation Results}
The system is evolved for a sufficient duration to ensure it reaches a stationary state.
The final configuration provides a direct point of comparison with the stationary solution of the VFP equation.

\subsubsection{Spatial Distribution and Wealth Inequality}
\begin{figure*}[ht]
\includegraphics[width=\textwidth]{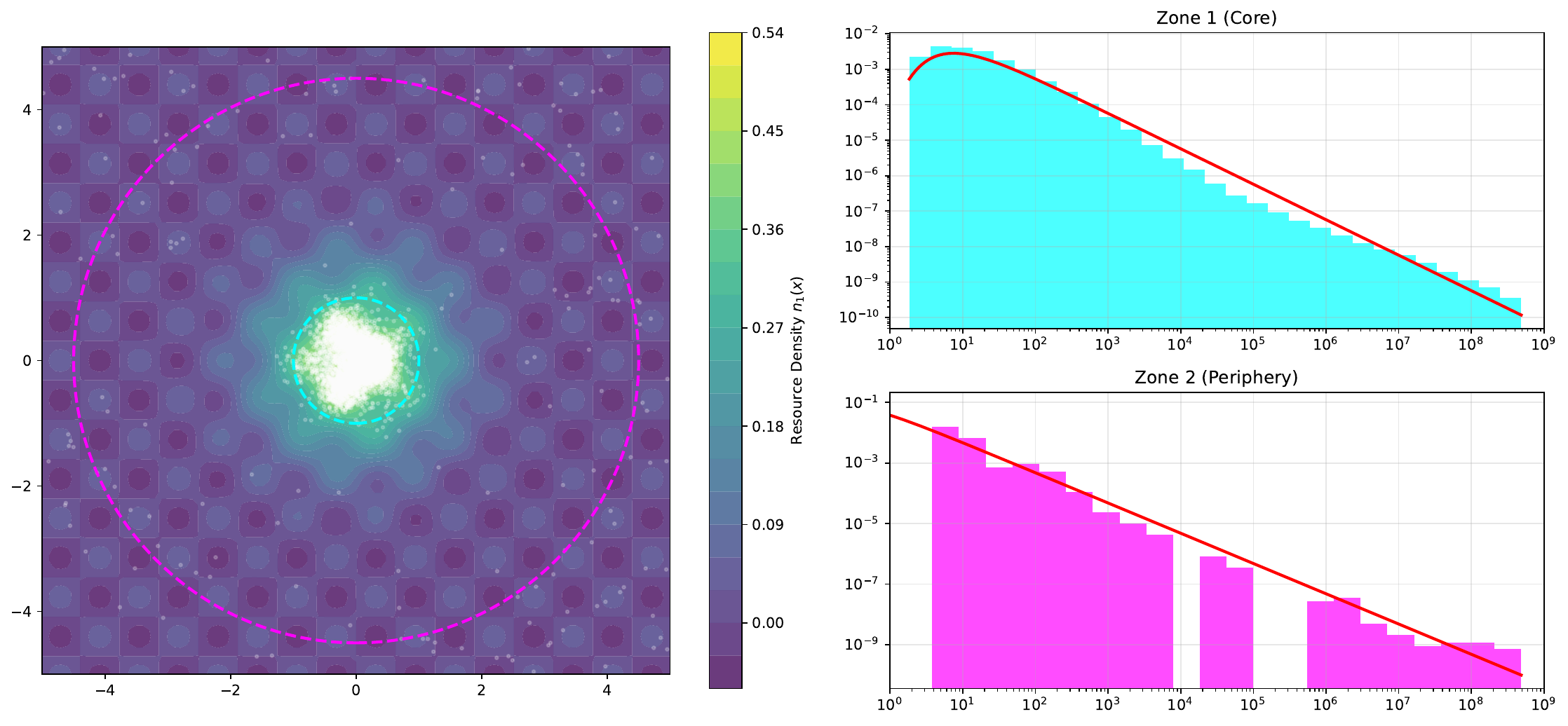}
\vfill
\includegraphics[width=\textwidth]{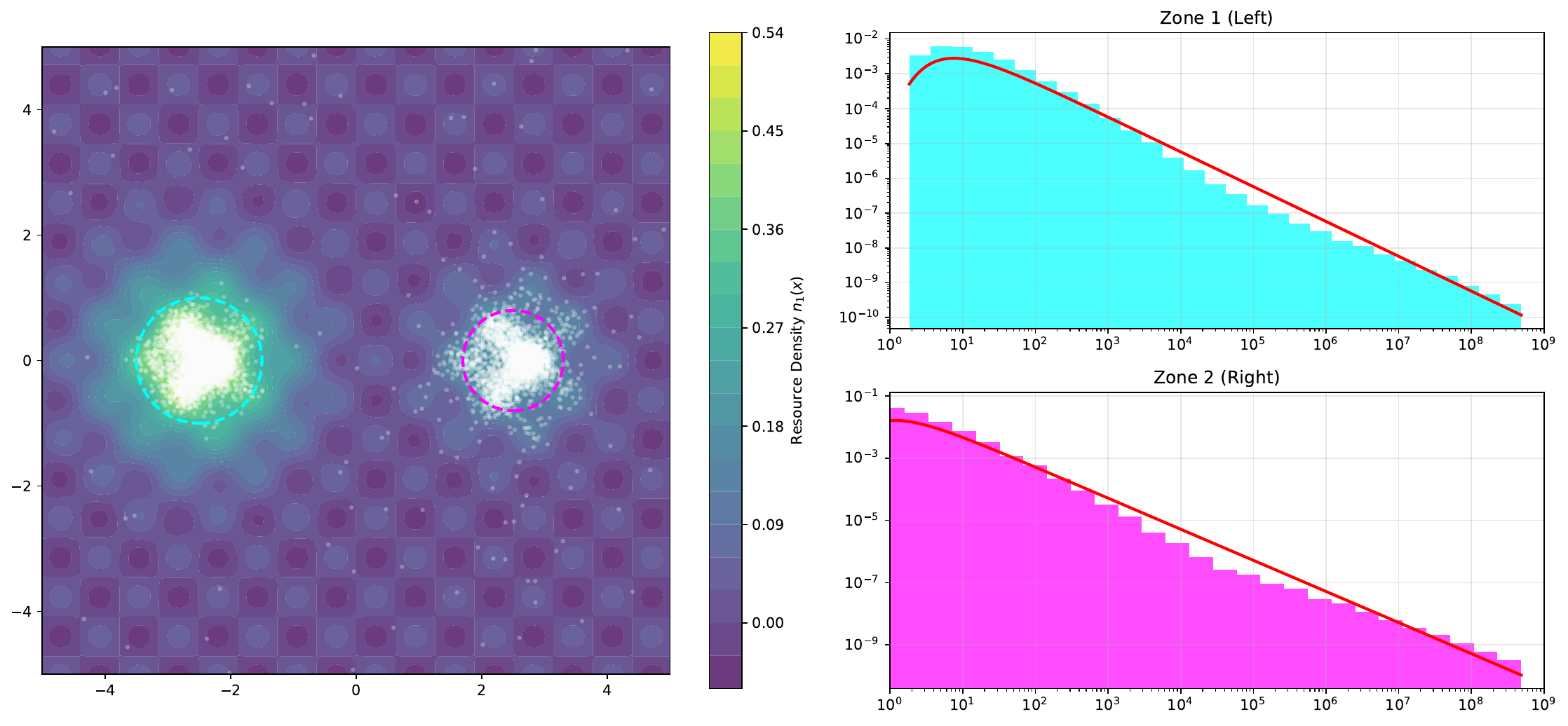}
\caption{\label{fig:distributions} Numerical validation of stationary state predictions for two different resource landscapes. \textbf{Top Row (Single-Well Landscape):} (a) The final spatial distribution of $L=10^5$ agents overlaid on a monocentric resource landscape. Agents concentrate in the central resource well. The ``Core'' (Zone 1, cyan) and ``Periphery'' (Zone 2, magenta) are defined for analysis. (b) Log-log plot of the wealth distribution for the agents in the core. (c) Log-log plot for the agents in the periphery. \textbf{Bottom Row (Double-Well Landscape):} (d) Final spatial distribution of agents in a polycentric landscape, with agents clustering in two distinct wells of different resource intensity. (e) Wealth distribution for the agents in the "Rich Well" (cyan). (f) Wealth distribution for the agents in the "Medium Well" (magenta). The histograms of simulated data (color bars) shows excellent agreement with the theoretical prediction (red solid lines).}
\end{figure*}

The results for the single-well and double-well landscapes are presented in Fig.~\ref{fig:distributions}. 
The simulations confirm that agents cluster in resource-rich areas and that the local wealth distributions match the theoretical prediction $\Psi(x,q)$ with remarkable accuracy. 
This validates the core prediction of our framework: the direct coupling of spatial and economic dynamics leads to emergent, spatially-dependent inequality. 
The agreement is not merely qualitative.
The theory correctly predicts not only the power-law tail of the wealth distribution but also the subtle shifts in the entire distribution's shape based on the local resource income $\mathcal{H}_{\text{eff}}(x)$.
This demonstrates that the quasi-stationary approximation made in Sec.~\ref{sec:application} is a valid and powerful tool for understanding the system's equilibrium.

\subsubsection{Quantitative Measures of Inequality}
\begin{figure*}[ht]
\includegraphics[width=0.48\textwidth]{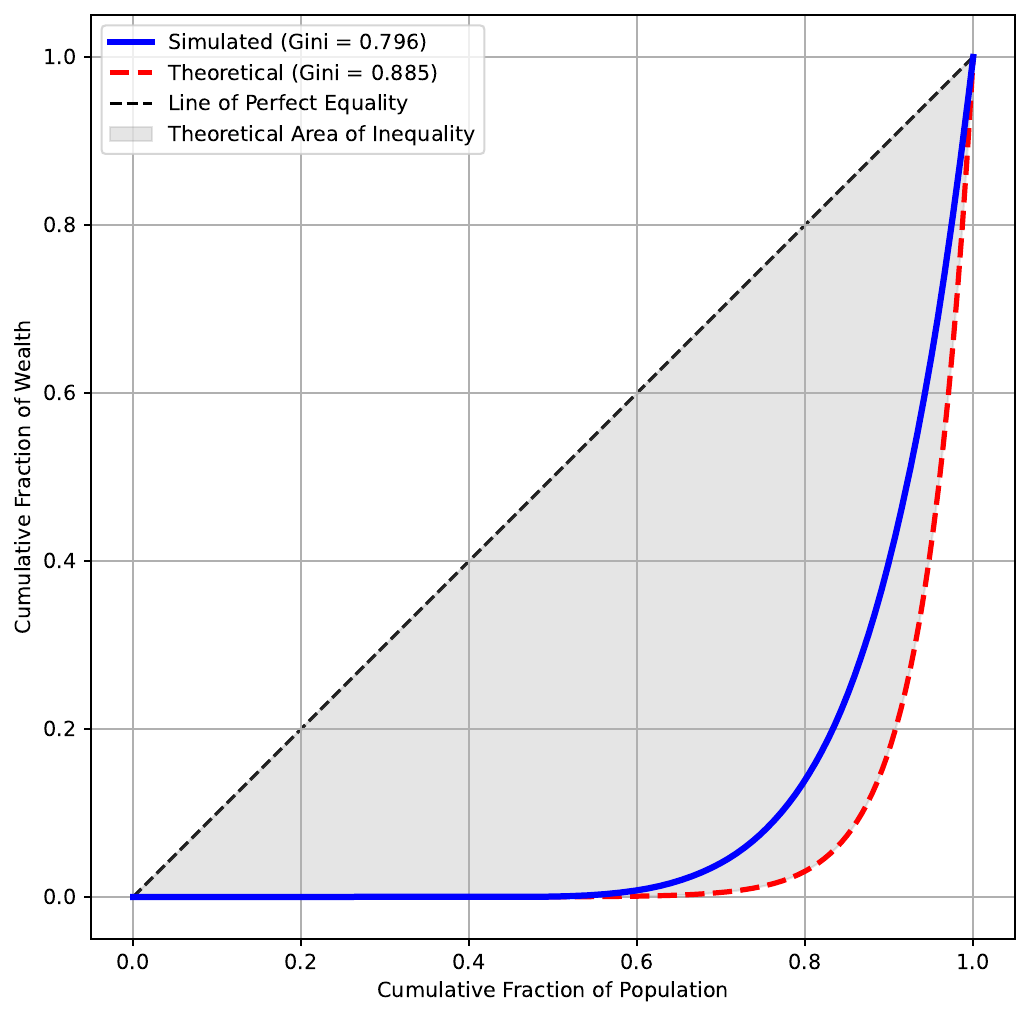}
\hfill
\includegraphics[width=0.48\textwidth]{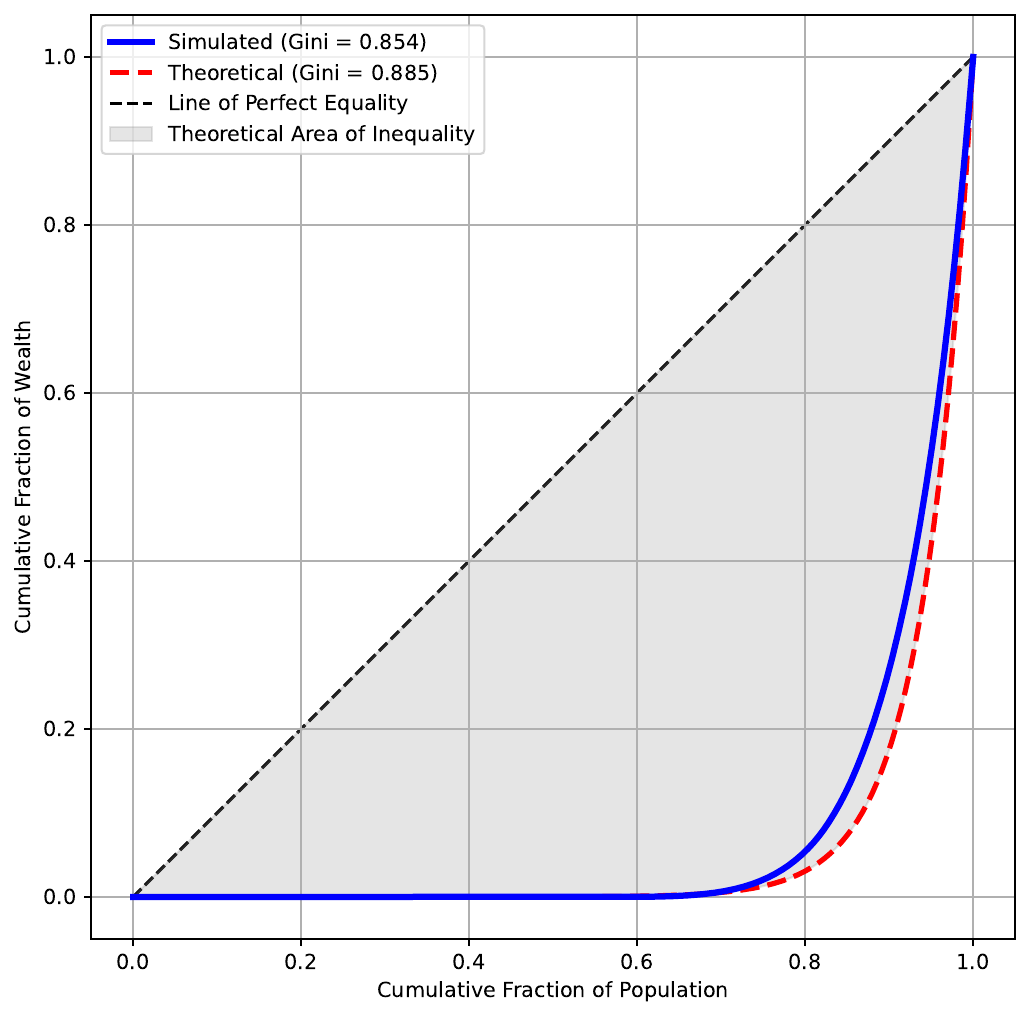}
\caption{\label{fig:lorenz_curves} Lorenz curves and Gini coefficients for the two simulated scenarios, compared with the analytical prediction. (a) \textbf{Single-Well Landscape:} The simulated data (solid teal line, $G \approx 0.796$) closely follows the theoretical Lorenz curve (dashed red line, $G \approx 0.885$). (b) \textbf{Double-Well Landscape:} The simulated Gini coefficient increases to $G \approx 0.854$, again tracking the theoretical prediction and demonstrating that greater landscape heterogeneity drives higher overall wealth inequality.}
\end{figure*}

Figure~\ref{fig:lorenz_curves} provides a quantitative test of the system's overall inequality structure by comparing the Lorenz curve from the simulated data to our analytical prediction.
In both the single-well and double-well scenarios, the simulated Lorenz curve shows a high degree of inequality, closely tracking the theoretical curve.

For the single-well case, the simulation yields a Gini coefficient of $G_{\text{sim}} \approx 0.796$, while the theory predicts $G_{\text{theory}} \approx 0.885$. 
In the double-well case, the simulated Gini is $G_{\text{sim}} \approx 0.854$.
The small discrepancy between the simulation and the mean-field theory is expected and physically meaningful.
This reflects the finite-size effects present in the simulation of $L=10^5$ agents, which are neglected in the $L \to \infty$ limit of the VFP theory.

The key result of this analysis and simulations is that the introduction of a more heterogeneous polycentric resource landscape leads to a measurable increase in the simulated Gini coefficient compared to the monocentric case. 
This confirms that the spatial structure of the environment is a direct driver of the magnitude of socio-economic inequality. 
This finding resonates with extensive research in economic geography and urban economics, which has long established a strong correlation between geographic factors, resource availability, and the concentration of economic activity and wealth~\cite{Krugman1991,Gallup1999}. 
Our work provides a microscopic explanation for this empirically observed phenomenon, showing how the interplay between spatial attraction and spatially-dependent income opportunities naturally leads to greater overall inequality in a more structured landscape.

\section{Discussion}
\label{sec:discussion}
The framework developed herein provides a robust bridge between the microscopic rules governing individual agent behavior and the macroscopic patterns that emerge in complex socio-economic systems.
By progressing systematically from microscopic SDEs to mesoscopic fluctuating equations, and finally to macroscopic kinetic equations, we have constructed a multi-scale theoretical lens through which to view these phenomena. 
We now discuss the interpretation of these results, the crucial role of fluctuations, and the limitations and future directions of our model.

\subsection{Interpretation of the VFP System and its Parameters}
The VFP system captures the deterministic evolution of the phase-space densities. 
The application to a static resource landscape demonstrates its power, showing how spatial concentrations of population and wealth emerge naturally as a stationary state of the dynamics. 
The structure of the VFP equation reveals a rich interplay of competing effects. 
The \emph{Vlasov} part of the operator describes a collisionless, deterministic flow of the density through phase space, driven by the mean-field potential. 
This is the part of the dynamics responsible for collective organized phenomena. 
In contrast, the \emph{Fokker-Planck} part describes the dissipative effects of a ``thermal bath,'' causing friction and diffusion in momentum space and driving the system towards a state of maximum entropy.

In a socio-economic context, the parameters of this system take on specific meanings. 
The ``social temperature'' $T_0$ is a particularly important parameter. 
It does not correspond to a physical temperature but rather quantifies the magnitude of idiosyncratic randomness in agent decision-making.
A low $T_0$ describes a system of near-optimizing agents who closely follow the gradients of the effective social and economic potentials, while a high $T_0$ represents a system where behavior is dominated by random exploration, aligning with concepts of bounded rationality~\cite{Simon1957}. 
The friction coefficient $\gamma_0$ can be interpreted as a measure of social inertia or resistance to change. 
The inclusion of inertia via the mass parameter $m_0$ is a novel feature for most social models, potentially relevant for describing overshoot phenomena in migrations or the damped oscillations of a population responding to a sudden economic shock.

\subsection{The Role of Fluctuations}

The underlying fluctuating kinetic equations provide a more complete and physically grounded picture than the mean-field limit alone. 
The DK system is exact for any finite number of agents and correctly describes the inherent stochasticity of the system. 
These fluctuations are not mere corrections to the mean-field behavior, but they are often the physical mechanism that drives the system's evolution, particularly for barrier-crossing phenomena.

For instance, the VFP equations may admit multiple stable stationary solutions (e.g., a spatially-uniform integrated state and a spatially-segregated state). 
The deterministic VFP dynamics cannot describe a transition from a metastable state to the true ground state.
It is the noise term in the DK equation, which scales as $1/\sqrt{L}$, provides the mechanism for such transitions. 
A random local fluctuation in density, which is naturally present in any system with a finite number of agents, can act as the ``seed'' for the nucleation of a new and more stable phase~\cite{Lutsko2012,LutskoDuranOlivencia2013CNTdynamical,LutskoDuranOlivencia2015CNTextension,DuranOlivenciaLutsko2015MNTconfined,duran_olivencia2018}, such as the spontaneous formation of a new settlement in a resource-rich area. 
While the mean-field equations describe the possible equilibrium landscapes, the fluctuating theory describes the stochastic pathways by which the system explores this landscape and settles into its preferred state.

\subsection{Model Assumptions, Limitations, and Future Directions}
The power of this framework also illuminates its limitations and points toward avenues for future research. 
The derivation relies on several key assumptions.
The mean-field limit, while powerful, neglects correlations between agents and is most accurate for systems with long-range interactions. 
The Maxwell-Boltzmann ansatz used in our application is a quasi-stationary approximation, valid under a separation of timescales between the fast physical dynamics and the slow economic dynamics.

Future work could proceed in several directions. 
First, the model could be extended to include demographic changes by incorporating birth-death processes, which would add non-linear reaction terms to the VFP equations. 
Second, the assumption of pairwise interactions could be relaxed. 
Many social and economic interactions are better described by networks~\cite{Newman2010}. 
Incorporating a network structure, where the interaction potentials $V_{ij}$ depend on a predefined adjacency matrix, would be a significant step towards greater realism. 
Finally, the simplified modeling of resource transport could be improved.
While our model captures local interactions, many processes, like trade or long-distance migration, are better described as non-local jumps. 
Replacing the Brownian motion in the microscopic SDEs with a Lévy process would lead to a kinetic theory with fractional derivatives, a challenging but potentially very rewarding direction for modeling human mobility~\cite{Brockmann2006}.
\section{Conclusion}
\label{sec:conclusion}
We have constructed a first-principles framework for a kinetic theory of multi-agent socio-economic systems. 
Starting from a microscopic description of agents governed by Langevin dynamics with coupled physical and economic state variables, we have systematically derived the exact mesoscopic dynamics in the form of a coupled system of DK equations.
We then formally derived the deterministic macroscopic dynamics, which take the form of a coupled VFP system, representing the mean-field limit of the theory.

The resulting framework provides a rigorous and versatile foundation for studying the emergence of complex social and economic patterns. 
We have demonstrated its utility by deriving a direct and analytical link between a non-uniform resource landscape and the emergence of correlated spatial and economic inequality. 
This application showcases the capacity of our model to connect microscopic agent rules to macroscopic and empirically relevant social structures—specifically, the spatial concentration of populations and the spontaneous formation of power-law wealth distributions. 
This work paves the way for more quantitative and predictive models of social dynamics, grounding the study of collective human behavior in the formal apparatus of non-equilibrium statistical physics.

\begin{acknowledgments}
The author is grateful to Antonio Malpica-Morales for formative discussions that helped spark the initial idea for this research. 
The author acknowledges financial support from Vortico Tech S.L.
The author also thanks the company for its encouragement and for fostering a supportive environment for this research project. 
The author is especially grateful to have been able to complete this work while accompanying his son during a hospital stay.
\end{acknowledgments}
\appendix

\section{Detailed Itô Calculation}
\label{app:ito}
This appendix provides a detailed application of the multidimensional Itô's Lemma, which forms the basis for the derivation in Sec.~\ref{sec:derivation}. Let $Z_t$ be a vector-valued stochastic process satisfying the SDE $dZ_t = \mathbf{b}(Z_t) dt + \mathbf{\Sigma}(Z_t) d\mathbf{W}_t$, where $\mathbf{b}$ is the drift vector and $\mathbf{\Sigma}$ is the diffusion matrix. For a scalar function $\phi(Z_t) \in C^2$, Itô's Lemma states~\cite{Oksendal2003}:
\begin{equation}
d\phi(Z_t) = \left( \mathbf{b} \cdot \nabla_Z \phi \right) dt + \frac{1}{2} \text{Tr}\left( \mathbf{\Sigma} \mathbf{\Sigma}^T D^2\phi \right) dt + \nabla_Z \phi \cdot \mathbf{\Sigma} d\mathbf{W}_t,
\end{equation}
where $D^2\phi$ is the Hessian matrix of $\phi$.

\subsection{Human Agents (k=0)}
For a human agent $i \in \mathcal{I}_0$, the state vector is $Z_i = (X_i, P_i, q_i)$, a $(2d+1)$-dimensional vector. The SDE system from Eqs.~(\ref{eq:sde_x}), (\ref{eq:sde_p}), and (\ref{eq:sde_q}) gives the drift vector $\mathbf{b}_i(Z)$ and diffusion matrix $\mathbf{\Sigma}_i(Z)$:
\begin{equation}
\mathbf{b}_i = 
\begin{pmatrix}
P_i/m_0 \\
F_i^{\text{int}}(Z) + F_i^{\text{ext}}(Z_i) - \gamma_0 P_i \\
f_i(Z_i, Z)
\end{pmatrix},
\end{equation}
\begin{equation}
\mathbf{\Sigma}_i = 
\begin{pmatrix}
\mathbf{0}_{d \times d} & \mathbf{0}_{d \times 1} \\
\sigma_0 \mathbf{I}_{d \times d} & \mathbf{0}_{d \times 1} \\
\mathbf{0}_{1 \times d} & g_i(Z_i, Z)
\end{pmatrix}.
\end{equation}
The quadratic variation term is $\mathbf{\Sigma}_i \mathbf{\Sigma}_i^T = \text{diag}(\mathbf{0}_{d \times d}, \sigma_0^2 \mathbf{I}_{d \times d}, g_i^2)$. The test function is $\phi_0(x, p, q)$ with gradient $\nabla_Z \phi_0 = (\nabla_x \phi_0, \nabla_p \phi_0, \partial_q \phi_0)^T$.

The first term in Itô's Lemma is the dot product of the drift and the gradient:
\begin{align}
\mathbf{b}_i \cdot \nabla_Z \phi_0 &= \frac{P_i}{m_0} \cdot \nabla_x \phi_0 \nonumber \\ 
&+ \left( F_i^{\text{int}} + F_i^{\text{ext}} - \gamma_0 P_i \right) \cdot \nabla_p \phi_0 \nonumber\\
&+ f_i \frac{\partial \phi_0}{\partial q}.
\end{align}
The second (trace) term becomes:
\begin{equation}
\frac{1}{2} \text{Tr}\left( \mathbf{\Sigma}_i \mathbf{\Sigma}_i^T D^2\phi_0 \right) = \frac{1}{2} \left( \sigma_0^2 \Delta_p \phi_0 + g_i^2 \frac{\partial^2 \phi_0}{\partial q^2} \right).
\end{equation}
The final stochastic term is:
\begin{equation}
\nabla_Z \phi_0 \cdot \mathbf{\Sigma}_i d\mathbf{W}_t = \nabla_p \phi_0 \cdot \sigma_0 dW_i(t) + \frac{\partial \phi_0}{\partial q} g_i dW'_i(t).
\end{equation}
Combining these terms yields the full stochastic differential for $\phi_0(Z_i)$.

\section{Explicit Derivation of the Generator and Martingale Terms}
\label{app:generator}
The generator $\mathcal{L}_k$ and martingale $dM_k^L$ are constructed from the terms of the Itô expansion for a single particle $i$ of type $k$. We detail the derivation for each agent type below.

\subsection{Human Agents (k=0)}
The Itô expansion for a test function $\phi_0(Z_i)$ applied to a single human agent $i \in \mathcal{I}_0$ is given by (see Appendix~\ref{app:ito}):
\begin{align}
d\phi_0(Z_i) =& \left[
\begin{array}{l}
 \frac{P_i}{m_0} \cdot \nabla_x \phi_0 \\
 \quad+ \left( F_i^{\text{ext}} + F_i^{\text{int}} - \gamma_0 P_i \right) \cdot \nabla_p \phi_0 \\
 \quad\quad+ f_i \frac{\partial \phi_0}{\partial q} + \frac{\sigma_0^2}{2} \Delta_p \phi_0 + \frac{g_i^2}{2} \frac{\partial^2 \phi_0}{\partial q^2} 
 \end{array}\right] dt \nonumber \\
&+ \sigma_0 \nabla_p \phi_0(Z_i) dW_i + g_i \frac{\partial \phi_0}{\partial q}(Z_i) dW'_i.
\end{align}
The generator $\mathcal{L}_0[\{\mu_j^L\}]\phi_0(z)$ is the operator representing the term in the square brackets, evaluated at a generic state $z=(x,p,q)$. To express it as a functional of the measures, we replace the interaction sums with their integral forms. For example, the interaction force on an agent at state $z$ is expressed as:
\begin{equation}
F_0^{\text{int}}(z, \{\mu_j^L\}) = L \sum_{k'=0}^{M} \int_{S_{k'}} F_{0,k'}^{\text{int}}(z,z') d\mu_{k'}^L(z').
\end{equation}
Applying this to all interaction terms gives the full generator as presented in Sec.~\ref{sec:derivation}. The martingale $dM_0^L(\phi_0)$ is obtained by summing the stochastic terms over all human agents and normalizing by $L$:
\begin{align}
dM_0^L(\phi_0) = \frac{1}{L} \sum_{i \in \mathcal{I}_0} \left( \sigma_0 \nabla_p \phi_0(Z_i) \cdot dW_i + g_i \frac{\partial \phi_0}{\partial q}(Z_i) dW'_i \right).
\end{align}

\subsection{Resource Agents (k $\ge$ 1)}
The procedure for resource agents is analogous but simpler due to the absence of the wealth variable. The Itô expansion for a test function $\phi_k(Z_j)$ applied to a single resource agent $j \in \mathcal{I}_k$ is:
\begin{align}
d\phi_k(Z_j) = &\left[ 
\begin{array}{l}
	\frac{P_j}{m_k} \cdot \nabla_x \phi_k \\
	\quad+ \left( F_j^{\text{ext}} + F_j^{\text{int}} - \gamma_k P_j \right) \cdot \nabla_p \phi_k \\
	\quad\quad+ \frac{\sigma_k^2}{2} \Delta_p \phi_k
\end{array} \right] dt \nonumber\\
&+ \sigma_k \nabla_p \phi_k(Z_j)  dW_j.
\end{align}
The generator $\mathcal{L}_k[\{\mu_j^L\}]\phi_k(z)$ is the operator in the square brackets, with the interaction forces written in terms of the empirical measures:
\begin{equation}
F_k^{\text{int}}(z, \{\mu_j^L\}) = L \sum_{k'=0}^{M} \int_{S_{k'}} F_{k,k'}^{\text{int}}(z,z') d\mu_{k'}^L(z').
\end{equation}
The martingale $dM_k^L(\phi_k)$ is the normalized sum of the stochastic terms:
\begin{align}
dM_k^L(\phi_k) = \frac{1}{L} \sum_{j \in \mathcal{I}_k} \sigma_k \nabla_p \phi_k(Z_j)  dW_j.
\end{align}
This completes the explicit derivation for both agent types.

\section{Equivalence of Energy Functional Representations}
\label{app:functional_equivalence}
In this appendix, we demonstrate the mathematical equivalence between the energy functionals expressed in terms of the phase-space empirical measures ($\mathcal{E}_K, \mathcal{E}_U$) and those expressed in terms of the microscopic configuration-space fields ($\mathcal{H}_K, \mathcal{H}_U$). 
Both represent the same underlying microscopic energy of the discrete particle system.

\subsection{Kinetic Energy}
The kinetic energy functional in terms of the phase-space measure is the direct definition of the total kinetic energy:
\begin{align}
    \mathcal{E}_K[\{\mu_j^L\}] &= L \sum_{k=0}^{M} \int_{S_k} \frac{|p|^2}{2m_k} d\mu_k^L(z) \nonumber \\
    &= L \sum_{k=0}^{M} \int_{S_k} \frac{|p|^2}{2m_k} \left(\frac{1}{L} \sum_{i \in \mathcal{I}_k} \delta(z-Z_i) \right) dz \nonumber \\
    &= \sum_{k=0}^{M} \sum_{i \in \mathcal{I}_k} \frac{|P_i|^2}{2m_k}.
\end{align}
The field-theoretic functional is $\mathcal{H}_K = \frac{1}{2} \int d\mathbf{r} \, \pi(\mathbf{r})^2 / \phi(\mathbf{r})$. Substituting the microscopic definitions $\phi(\mathbf{r}) = \sum_i \delta(\mathbf{r}-\mathbf{r}_i)$ and $\pi(\mathbf{r}) = \sum_i \mathbf{p}_i \delta(\mathbf{r}-\mathbf{r}_i)$, the integrand can be understood in a discretized sense, as discussed in~\cite{duran_olivencia2017}. In any small volume element containing only particle $j$, the integrand becomes $(\mathbf{p}_j^2 \delta_j^2)/\delta_j = \mathbf{p}_j^2 \delta_j$. Integrating over all space then yields $\sum_j |\mathbf{p}_j|^2$. Thus, we have:
\begin{equation}
    \mathcal{H}_K = \frac{1}{2} \sum_{j} |\mathbf{p}_j|^2 = \mathcal{E}_K.
\end{equation}

\subsection{Potential Energy}
The interaction part of the potential energy functional in terms of the empirical measures is:
\begin{align}
    \mathcal{E}_{\text{int}}[\{\mu_j^L\}] &= \frac{L^2}{2} \sum_{k,k'} \iint V_{k,k'}(z,z') d\mu_k^L(z) d\mu_{k'}^L(z') \nonumber \\
    &= \frac{1}{2} \sum_{k,k'} \sum_{i \in \mathcal{I}_k} \sum_{j \in \mathcal{I}_{k'}} V_{k,k'}(Z_i, Z_j).
\end{align}
This formulation represents the sum over all pairs, including self-interaction terms where $i=j$.

The field-theoretic functional is:
\begin{equation}
	\mathcal{H}_U = \frac{1}{2} \iint d\mathbf{r}d\mathbf{r}' V(\mathbf{r}-\mathbf{r}')(\phi(\mathbf{r})\phi(\mathbf{r}') - \delta(\mathbf{r}-\mathbf{r}')\phi(\mathbf{r})).
	\end{equation} 
Substituting the microscopic density field $\phi(\mathbf{r}) = \sum_i \delta(\mathbf{r}-\mathbf{r}_i)$:
\begin{itemize}
    \item The first term, $\iint V \phi\phi'$, becomes $\frac{1}{2} \sum_{i,j} V(\mathbf{r}_i - \mathbf{r}_j)$.
    \item The second term, $-\frac{1}{2} \iint V \delta(\mathbf{r}-\mathbf{r}')\phi(\mathbf{r})$, becomes $-\frac{1}{2} \int V(0) \phi(\mathbf{r}) d\mathbf{r} = -\frac{1}{2} \sum_i V(0)$.
\end{itemize}
This term explicitly subtracts the $i=j$ (self-interaction) contributions. Thus,
\begin{equation}
    \mathcal{H}_U = \frac{1}{2} \sum_{i \neq j} V(\mathbf{r}_i - \mathbf{r}_j).
\end{equation}
The functionals are therefore equivalent up to the treatment of self-interaction. For deriving the interaction force on particle $i$ from particles $j \neq i$, both functionals yield the same result via their respective functional derivatives.

\section{From the Weak Form to the SPDE}
\label{app:adjoint}
This appendix details the formal derivation of the SPDE for $\mu_0^L$ from its weak-form evolution equation. We start with the deterministic part of Eq.~(\ref{eq:sde_mu0}), which equates the time evolution of the average of a test function with the average of the generator acting on that test function:
\begin{equation}
    \frac{d}{dt} \langle \mu_0^L, \phi_0 \rangle = \langle \mu_0^L, \mathcal{L}_0 \phi_0 \rangle.
\end{equation}
Writing out the inner products as integrals over the phase space $S_0$ gives:
\begin{equation}
    \int_{S_0} \frac{\partial \mu_0^L}{\partial t} \phi_0 dz = \int_{S_0} \mu_0^L (\mathcal{L}_0 \phi_0) dz.
\end{equation}
To find the strong-form equation for $\mu_0^L$, we apply integration by parts to each term of the integral on the right-hand side. This procedure formally moves all differential operators from the test function $\phi_0$ to the measure $\mu_0^L$. We assume that $\phi_0$ has compact support, ensuring that all boundary terms resulting from the integration by parts are zero. Each term of $\int \mu_0^L (\mathcal{L}_0 \phi_0) dz$ is transformed as follows:
$$ \int \mu_0^L \left( \frac{p}{m_0} \cdot \nabla_x \phi_0 \right) dz = - \int \phi_0 \left( \nabla_x \cdot \left(\frac{p}{m_0} \mu_0^L\right) \right) dz $$
$$ \int \mu_0^L \left( \mathbf{F}_{\text{total}} \cdot \nabla_p \phi_0 \right) dz = - \int \phi_0 \left( \nabla_p \cdot \left( \mathbf{F}_{\text{total}} \mu_0^L \right) \right) dz $$
$$ \int \mu_0^L \left( \frac{\sigma_0^2}{2} \Delta_p \phi_0 \right) dz = \int \phi_0 \left( \frac{\sigma_0^2}{2} \Delta_p \mu_0^L \right) dz $$
$$ \int \mu_0^L \left( f_{\text{total}} \frac{\partial \phi_0}{\partial q} \right) dz = - \int \phi_0 \left( \frac{\partial}{\partial q} \left(f_{\text{total}} \mu_0^L\right) \right) dz $$
$$ \int \mu_0^L \left( \frac{g_0^2}{2} \frac{\partial^2 \phi_0}{\partial q^2} \right) dz = \int \phi_0 \left( \frac{1}{2} \frac{\partial^2}{\partial q^2} \left(g_0^2 \mu_0^L\right) \right) dz $$
where $\mathbf{F}_{\text{total}}$ and $f_{\text{total}}$ are the total drift terms in momentum and wealth, respectively.

By substituting these transformed terms back into the integral equation, we find that all differential operators now act on $\mu_0^L$. Since the resulting equality must hold for any arbitrary test function $\phi_0$, we can equate the integrands. This allows us to identify the deterministic part of the SPDE, which is governed by the formal adjoint operator $\mathcal{L}_0^*$:
\begin{align}
\frac{\partial \mu_0^L}{\partial t} =&\ \mathcal{L}_0^* \mu_0^L \nonumber\\
=& - \nabla_x \cdot \left(\frac{p}{m_0} \mu_0^L\right) + \nabla_p \cdot \left( -\mathbf{F}_{\text{total}} \mu_0^L \right) \nonumber \\
& + \frac{\sigma_0^2}{2} \Delta_p \mu_0^L- \frac{\partial}{\partial q} \left(f_{\text{total}} \mu_0^L\right) + \frac{1}{2} \frac{\partial^2}{\partial q^2} \left(g_0^2 \mu_0^L\right).
\end{align}
The final step is to formally associate the martingale term $dM_0^L$ with the stochastic noise terms in the final SPDE. This is achieved by constructing stochastic fluxes in the strong-form equation whose weak-form integrals are martingales that possess the same quadratic variation as $dM_0^L$. The explicit proof that the noise terms in Eq.~(\ref{eq:dk_human}) satisfy this requirement is provided in Appendix~\ref{app:noise_justification}.

\section{Justification of the Fluctuation Term in the SPDE}
\label{app:noise_justification}
This appendix justifies the form of the noise term in the fluctuating kinetic equations. The form of the SPDE must be consistent with the underlying particle dynamics, meaning the quadratic variation of its weak-form solution must match that of the martingale derived from the particle system.

\subsection{Resource Agents (k $\ge$ 1)}
The martingale for resource agents is:
\begin{equation}
	dM_k^L(\phi_k) = \frac{1}{L} \sum_{j \in \mathcal{I}_k} \sigma_k \nabla_p \phi_k(Z_j) \cdot dW_j(t).
\end{equation}
Its quadratic variation is:
\begin{align}
d[M_k^L(\phi_k)]_t =& \frac{\sigma_k^2}{L^2} \sum_{j \in \mathcal{I}_k} |\nabla_p \phi_k(Z_j)|^2 dt \nonumber\\
=& \frac{\sigma_k^2}{L} \langle \mu_k^L, |\nabla_p \phi_k|^2 \rangle dt.
\end{align}
Now, consider the proposed noise term in the SPDE, $d\mathcal{N}_k^L = \nabla_p \cdot (\mathbf{B}_k \circ d\mathbf{W}_p^{(k)})$. Its contribution to the average is $\langle d\mathcal{N}_k^L, \phi_k \rangle = - \langle \mathbf{B}_k \circ d\mathbf{W}_p^{(k)}, \nabla_p \phi_k \rangle$. The quadratic variation of this stochastic integral is:
$$
d[\langle d\mathcal{N}_k^L, \phi_k \rangle]_t = \left( \int_{S_k} (\nabla_p \phi_k)^T \mathbf{B}_k \mathbf{B}_k^T (\nabla_p \phi_k) d(x,p) \right) dt.
$$
Equating the two quadratic variations requires that the operator $\mathbf{B}_k \mathbf{B}_k^T$ be a measure itself, such that for any test function $\psi$, $\int \psi \mathbf{B}_k \mathbf{B}_k^T dz = \frac{\sigma_k^2}{L} \langle \mu_k^L, \psi \rangle$. This is satisfied by choosing:
$$
\mathbf{B}_k \mathbf{B}_k^T = \frac{\sigma_k^2}{L} \mu_k^L \mathbf{I}_{d \times d}.
$$
This justifies the form $\mathbf{B}_k = \sqrt{\frac{\sigma_k^2}{L} \mu_k^L} \mathbf{I}_{d \times d}$ for the noise amplitude in the SPDE.

\subsection{Human Agents (k=0)}
The martingale for human agents contains two independent Wiener processes, one for momentum ($dW_i$) and one for wealth ($dW'_i$):
$$
dM_0^L(\phi_0) = \frac{1}{L} \sum_{i \in \mathcal{I}_0} \left( \sigma_0 \nabla_p \phi_0(Z_i) \cdot dW_i + g_i \frac{\partial \phi_0}{\partial q}(Z_i) dW'_i \right).
$$
Because the Wiener processes are independent, the quadratic variation is the sum of the individual quadratic variations:
\begin{align*}
d[M_0^L(\phi_0)]_t &= \frac{1}{L^2} \sum_{i \in \mathcal{I}_0} \left( \sigma_0^2 |\nabla_p \phi_0(Z_i)|^2 + g_i^2 \left|\frac{\partial \phi_0}{\partial q}(Z_i)\right|^2 \right) dt \\
&= \frac{1}{L} \left\langle \mu_0^L, \sigma_0^2 |\nabla_p \phi_0|^2 + g_0^2 \left|\frac{\partial \phi_0}{\partial q}\right|^2 \right\rangle dt.
\end{align*}
The proposed noise term in the SPDE also has two independent parts: 
\begin{equation}
d\mathcal{N}_0^L = \nabla_p \cdot (\mathbf{B}_p \circ d\mathbf{W}_p) + \frac{\partial}{\partial q} (B_q \circ dW_q).
\end{equation} 
The quadratic variation of its weak form is likewise the sum of two terms:
$$
d[\langle d\mathcal{N}_0^L, \phi_0 \rangle]_t = \left(\begin{array}{l}
 \int_{S_0} |\nabla_p \phi_0|^2 \mathbf{B}_p \mathbf{B}_p^T dz\\
 \quad + \int_{S_0} \left|\frac{\partial \phi_0}{\partial q}\right|^2 B_q B_q^T dz 
 \end{array}\right) dt.
$$
Equating the momentum and wealth parts separately, we find:
$$
\mathbf{B}_p \mathbf{B}_p^T = \frac{\sigma_0^2}{L} \mu_0^L \mathbf{I}_{d \times d} \quad \text{and} \quad B_q B_q^T = \frac{g_0^2}{L} \mu_0^L.
$$
This justifies the form of both noise amplitudes in Eq.~(\ref{eq:dk_human}).

\section{Functional Derivative Connection}
\label{app:functional_derivation_connection}
This appendix provides the formal derivation connecting the microscopic potential energy functional $\mathcal{E}_U[\{\mu_j^L\}]$ and the mesoscopic excess free energy functional $\mathcal{F}_{\text{exc}}[n_k]$.

\subsubsection{The Functional Chain Rule}
The connection is established using the chain rule for functional derivatives. The potential energy $\mathcal{E}_U$ is a functional of the phase-space measure $\mu_k(x,p)$, which in a local-equilibrium state can be written in terms of the configuration-space density $n_k(x)$ and a conditional momentum distribution $f_k(p|x)$:
$$
\mu_k(x,p) = n_k(x) f_k(p|x), \quad \text{with} \quad \int f_k(p|x) dp = 1.
$$
The derivative of $\mathcal{E}_U$ with respect to $n_k$ is given by the chain rule:
\begin{equation}
    \frac{\delta \mathcal{E}_U}{\delta n_k(x_0)} = \iint \frac{\delta \mathcal{E}_U}{\delta \mu_k(x,p)} \frac{\delta \mu_k(x,p)}{\delta n_k(x_0)} dx dp.
\end{equation}
The inner derivative is $\frac{\delta \mu_k(x,p)}{\delta n_k(x_0)} = \delta(x-x_0) f_k(p|x)$. Substituting this in and evaluating the integral over $x$ yields:
\begin{equation}
    \frac{\delta \mathcal{E}_U}{\delta n_k(x_0)} = \int \left( \frac{\delta \mathcal{E}_U}{\delta \mu_k(x_0,p)} \right) f_k(p|x_0) dp.
    \label{eq:app_integral_form}
\end{equation}

\subsubsection{Evaluation of the Microscopic Derivative}
We now find the explicit form of the derivative $\frac{\delta \mathcal{E}_U}{\delta \mu_k(x_0,p)}$. From the definition of $\mathcal{E}_U$:
\begin{align}
    \mathcal{E}_U[\{\mu_j^L\}] =& L \sum_{j} \int U_j(z) d\mu_j^L(z) \nonumber \\
    & + \frac{L^2}{2} \sum_{j,j'} \iint V_{j,j'}(z,z') d\mu_j^L(z) d\mu_{j'}^L(z').
\end{align}
Taking the functional derivative with respect to $\mu_k$ at a specific phase-space point $z_0 = (x_0, p_0)$ gives:
\begin{equation}
    \frac{\delta \mathcal{E}_U}{\delta \mu_k(z_0)} = L U_k(z_0) + L^2 \sum_{k'=0}^{M} \int_{S_{k'}} V_{k,k'}(z_0,z') \mu_{k'}(z') dz'.
\end{equation}
The interaction potentials in our physical model, $U_k(z)$ and $V_{k,k'}(z,z')$, depend only on the spatial coordinates (and potentially wealth), but are \emph{independent of momentum}. Therefore, the functional derivative above is also independent of momentum, $p_0$. We can thus denote it as $U_k^{\text{eff}}(x_0, [\mu_j^L])$.

\subsubsection{Final Result and the Ensemble Average}
With the derivative being independent of momentum, we can now solve the integral in Eq.~(\ref{eq:app_integral_form}). We can pull the term $U_k^{\text{eff}}(x_0, [\mu_j^L])$ outside the momentum integral:
\begin{align}
    \frac{\delta \mathcal{E}_U}{\delta n_k(x_0)} &= \int U_k^{\text{eff}}(x_0, [\mu_j^L]) f_k(p|x_0) dp \nonumber \\
    &= U_k^{\text{eff}}(x_0, [\mu_j^L]) \int f_k(p|x_0) dp.
\end{align}
Since the conditional probability distribution for momentum must be normalized, $\int f_k(p|x_0) dp = 1$, we arrive at the result for a single microscopic configuration:
\begin{equation}
    \frac{\delta \mathcal{E}_U}{\delta n_k(x_0)} = U_k^{\text{eff}}(x_0, [\mu_j^L]).
\end{equation}
This result is for one fluctuating microstate. To obtain the mesoscopic thermodynamic free energy $\mathcal{F}_{\text{exc}}$, we must average over all microstates consistent with the given average density profile $n_k(x)$. This is the local equilibrium average $\langle \dots \rangle_{\text{leq}}$:
\begin{equation}
    \frac{\delta \mathcal{F}_{\text{exc}}[n_k]}{\delta n_k(x)} := \left\langle \frac{\delta \mathcal{E}_U[\{\mu_j^L\}]}{\delta n_k(x)} \right\rangle_{\text{leq}}.
\end{equation}
This provides the formal and explicit link between the microscopic potential energy and the mesoscopic excess free energy~\cite{duran_olivencia2017}.

\section{Details of the Mean-Field Limit}
\label{app:mean_field}
This appendix provides further details on the formal derivation of the Vlasov-Fokker-Planck system from the SDE for the empirical measure averages. The limit is taken as the number of particles $L \to \infty$.

\subsection{Scaling of Interaction Potentials}
A crucial assumption in mean-field theory is that the pairwise interaction strength scales inversely with the number of particles, such that the total force on any given agent remains finite. We thus posit the scaling $V_{k,k'} \to V_{k,k'}/L$ and $H_{0,k'} \to H_{0,k'}/L$. The interaction force on particle $i$ in the generator is then:
\begin{align}
F_i^{\text{int}}(Z) =&\ L \sum_{k'} \int F_{k_i,k'}^{\text{int}}(Z_i, z') d\mu_{k'}^L(z') \nonumber\\
=&\ \sum_{k'} \sum_{j \in \mathcal{I}_{k'}} \frac{1}{L} F_{k_i,k_j}^{\text{int}}(Z_i, Z_j).\nonumber
\end{align}
This scaling ensures a well-defined limit.

\subsection{Convergence of Terms in the Limit $L \to \infty$}
We examine the limit of the terms in the weak-form SDE, e.g., Eq.~(\ref{eq:sde_mu0}).

\subsubsection{Vanishing of the Martingale Term} 
The expected total quadratic variation of the martingale $M_k^L(\phi_k)$ over an interval $[0, T]$ is of order $\mathbb{E}[M_k^L(\phi_k)]_T \propto T/L$. As $L \to \infty$, this term vanishes, implying the martingale itself converges to zero in probability. Thus, the stochastic nature of the evolution disappears in the limit, leaving a deterministic equation.

\subsubsection{Convergence of the Interaction Term}
With the $1/L$ scaling, the average of the interaction force component in the generator becomes a U-statistic. By the law of large numbers for empirical measures and the propagation of chaos assumption, this double average converges to a double integral over the limiting deterministic measures. For the human agents, for example:
    \begin{align*}
    \lim_{L\to\infty} \left\langle \mu_0^L, F_{\text{int}} \cdot \nabla_p \phi_0 \right\rangle &= \\
    & \hspace{-3.5cm} \int_{S_0} \left( \sum_{k'=0}^M \int_{S_{k'}} F_{0,k'}^{\text{int}}(z,z') \rho_{k'}(t,z')dz' \right)  \nabla_p \phi_0(z) \rho_0(t,z)dz.
    \end{align*}
    This is the weak form of the Vlasov interaction term in Eq.~(\ref{eq:vfp_human_explicit}). The same logic applies to all interaction terms, justifying the transition from sums over empirical measures to integrals over the limiting densities.

\section{Cancellation of the VFP Operator with the Maxwell-Boltzmann Ansatz}
\label{app:boltzmann_cancellation}
In this appendix, we explicitly demonstrate that the Maxwell-Boltzmann ansatz
$$
\rho_0(x,p,q) = \Psi(x,q) \exp(-E(x,p)/k_B T_0),
$$
causes the spatial and momentum parts of the stationary VFP operator to vanish, provided the fluctuation-dissipation relation holds. The relevant part of the stationary VFP equation is:
\begin{equation}
    0 = -\nabla_x \cdot \left(\frac{p}{m_0} \rho_0\right) + \nabla_p \cdot \left( \gamma_0 p \rho_0 - \mathbf{F}_{\text{eff}}(x)\rho_0 \right) + \frac{\sigma_0^2}{2} \Delta_p \rho_0.
\end{equation}
We substitute the ansatz:
$$\rho_0 = \Psi(x,q) e^{-|p|^2/(2m_0 k_B T_0)} e^{-U_{\text{eff}}(x)/k_B T_0}$$ 
and analyze each term. 

The advection term becomes:
\begin{align}
    -\nabla_x \cdot \left(\frac{p}{m_0} \rho_0\right) = -\frac{p}{m_0} \cdot \nabla_x \rho_0 = \frac{\mathbf{p} \cdot \nabla_x U_{\text{eff}}(x)}{m_0 k_B T_0} \rho_0.
\end{align}
For the Vlasov (force) term, using $\mathbf{F}_{\text{eff}}(x) = -\nabla_x U_{\text{eff}}(x)$ and noting that it is independent of $p$, we have:
\begin{align}
    -\nabla_p \cdot (\mathbf{F}_{\text{eff}}(x) \rho_0) = -\mathbf{F}_{\text{eff}}(x) \cdot \nabla_p \rho_0 = \frac{\mathbf{F}_{\text{eff}}(x) \cdot \mathbf{p}}{m_0 k_B T_0} \rho_0.
\end{align}
This term is exactly $-\frac{\nabla_x U_{\text{eff}}(x) \cdot \mathbf{p}}{m_0 k_B T_0} \rho_0$ and thus precisely cancels the advection term.

For the Fokker-Planck (friction and diffusion) terms, we first expand the friction part:
\begin{align}
\nabla_p \cdot (\gamma_0 p \rho_0) =&\ (\nabla_p \cdot (\gamma_0 p)) \rho_0 + \gamma_0 p \cdot \nabla_p \rho_0\nonumber \\ 
=& d\gamma_0 \rho_0 - \frac{\gamma_0 |p|^2}{m_0 k_B T_0} \rho_0.\nonumber
\end{align}
Next, the Laplacian of $\rho_0$ with respect to $p$ is:
\begin{align}
\Delta_p \rho_0 =&\ \nabla_p \cdot \left(-\frac{p}{m_0 k_B T_0}\rho_0\right) \nonumber\\
=&\ \left( -\frac{d}{m_0 k_B T_0} + \frac{|p|^2}{(m_0 k_B T_0)^2} \right) \rho_0.
\end{align}
Combining these gives the full Fokker-Planck operator acting on $\rho_0$:
$$
\mathcal{L}_{\text{FP}}=\left[ \begin{array}{l}
\left(d\gamma_0 - \frac{d\sigma_0^2}{2m_0 k_B T_0}\right) \\
\quad + |p|^2\left(-\frac{\gamma_0}{m_0 k_B T_0} + \frac{\sigma_0^2}{2(m_0 k_B T_0)^2}\right)
\end{array}\right] \rho_0.
$$
Now, if we use the fluctuation-dissipation relation $\sigma_0^2 = 2\gamma_0 m_0 k_B T_0$, both terms in the square brackets are identically zero. Since the Vlasov and Fokker-Planck parts of the operator sum to zero when acting on the ansatz, only the terms involving derivatives in $q$ remain, as stated in the main text.
\bibliographystyle{unsrt}
\bibliography{references}
\end{document}